\documentclass{aa}  

\usepackage{graphicx}
\usepackage{color}

\usepackage{txfonts}
\usepackage{hyperref}
\newcommand{\gaia}{\textit{Gaia}}
\definecolor{dgreen}{rgb}{0.1, 0.53, 0.22}
 
\usepackage{lineno}
\begin{document}

   \title{Cepheid Metallicity in the Leavitt Law (C- MetaLL) survey: VI: Radial abundance gradients of 29 chemical species in the Milky Way disc \thanks{Based on the European Southern Observatory programs 108.227Z; 109.231T; 110.23WM and on the Telescopio Nazionale Galileo programmes A43TAC\_16; A44TAC\_27; A45TAC\_12; A46TAC\_15. Based on observations obtained at the Canada-France-Hawaii Telescope (CFHT) which is operated by the National Research Council of Canada, the Institut National des Sciences de l´Univers of the Centre National de la Recherche Scientique of France, and the University of Hawaii.}}
\titlerunning{C-MetaLL Survey VI}
   \author{E. Trentin\inst{1,2,3}\thanks{E-mail: etrentin@aip.de}
          \and
          G. Catanzaro\inst{4}
          \and 
          V. Ripepi\inst{3}
          \and
          J. Alonso-Santiago\inst{4}
          \and
          R. Molinaro\inst{3}
          \and
          J. Storm\inst{1}
          \and
          G. De Somma\inst{3,5,6}
          \and
          M. Marconi\inst{3} 
          \and
          A. Bhardwaj\inst{7}
          \and 
          M. Gatto\inst{3} 
          \and
          I. Musella \inst{3}
          \and
          V. Testa \inst{8}
          }

\institute{Leibniz-Institut f\"{u}r Astrophysik Potsdam (AIP), An der Sternwarte 16, D-14482 Potsdam, Germany
\and
Institut für Physik und Astronomie, Universität Potsdam, Haus 28, Karl-Liebknecht-Str. 24/25, D-14476 Golm (Potsdam), Germany
\and
INAF-Osservatorio Astronomico di Capodimonte, Salita Moiariello 16, 80131, Naples, Italy
\and
INAF-Osservatorio Astrofisico di Catania, Via S.Sofia 78, 95123, Catania, Italy 
\and
INAF-Osservatorio Astronomico d'Abruzzo, Via Maggini sn, 64100 Teramo, Italy
\and
Istituto Nazionale di Fisica Nucleare (INFN)-Sez. di Napoli, Via Cinthia, 80126 Napoli, Italy
\and
Inter-University Center for Astronomy and Astrophysics (IUCAA), Post Bag 4, Ganeshkhind, Pune 411 007, India
\and
INAF – Osservatorio Astronomico di Roma, via Frascati 33, I-00078 Monte Porzio Catone, Italy 
}
   \date{Received September 15, 1996; accepted March 16, 1997}

 
  \abstract
   {Classical Cepheids (DCEPs) are crucial for calibrating the extragalactic distance ladder, ultimately enabling the determination of the Hubble constant through the period-luminosity ($PL$) and period-Wesenheit ($PW$) relations that they exhibit. Hence, it is vital to understand how the $PL$ and $PW$ relations depend on metallicity. This is the purpose of the C-MetaLL survey, within which this work is situated. The DCEPs are also very important tracers of the young populations placed along the Galactic disc.}
   {We aim to enlarge the sample of DCEPs with accurate abundances from high-resolution spectroscopy. In particular, our goal is to extend the range of measured metallicities towards the metal-poor regime to better cover the parameter space. To this end, we observed objects in a wide range of Galactocentric radii, allowing us to study in detail the abundance gradients present in the Galactic disc. }
   {We present the results of the analysis of 331 spectra obtained for 180 individual DCEPs with a variety of high-resolution spectrographs. For each target, we derived accurate atmospheric parameters, radial velocities, and abundances for up to 29 different species. The iron abundances range between 0.5 and $-$1 dex with a rather homogeneous distribution in metallicity. }
   {The sample presented in this paper was complemented with that already published in the context of the C-MetaLL survey, resulting in a total of 292 pulsators whose spectra have been analysed in a homogeneous way. These data were used to study the abundance gradients of the Galactic disc in a range of Galactocentric radii ($R_{GC}$) spanning the range of 5-20 kpc.}
   {For most of the elements, we have found a clear negative gradient, with a slope of  $-0.064 \pm 0.003$ dex kpc$^{-1}$ for [Fe/H] case. Through a qualitative fit with the Galactic spiral arms, we show how our farthest targets ( $R_{GC}$ >10 kpc) trace both the Outer and Outer Scutum-Centaurus arms. The homogeneity of the sample will be of pivotal importance for the study of the metallicity dependence of the DCEP $PL$ relations.}

   \keywords{stars: abundances --
                stars: distances --
                stars: fundamental parameters --
                stars: variables: Cepheids --
                Galaxy: disc
               }

   \maketitle
%

\section{Introduction}

Classical Cepheids (DCEPs) play a fundamental role as standard candles in determining extragalactic distances due to the Leavitt law \citep{Leavitt1912}, a relationship between period and luminosity ($PL$). Calibrated using independent distances based on geometric methods such as trigonometric parallaxes, eclipsing binaries, and water masers, these relations serve as the first step in constructing the cosmic distance scale. This way, they calibrate secondary distance indicators, including type Ia supernovae (SNe Ia), allowing us to measure the distances of distant galaxies in the steady Hubble flow. The calibration of this three-step procedure, often referred to as the cosmic distance ladder, enables us to reach the Hubble flow and measure the Hubble constant, $H_0$, which connects the distance and the recession velocity of galaxies \citep[e.g.][and references therein]{Sandage2006,Freedman2012,Riess2016}.

\begin{sidewaystable}
\footnotesize\setlength{\tabcolsep}{3pt}
\caption{Main properties of the 180 programme DCEPs}
\label{tab:programStars}
  \begin{tabular}{lrllrlrrrrrrrll}
\hline
  \multicolumn{1}{c}{Star} &
  \multicolumn{1}{c}{Gaia\_source\_id} &
  \multicolumn{1}{c}{RA} &
  \multicolumn{1}{c}{Dec} &
  \multicolumn{1}{c}{Period} &
  \multicolumn{1}{c}{Mode} &
  \multicolumn{1}{c}{$G$} &
  \multicolumn{1}{c}{$G_{BP}$} &
  \multicolumn{1}{c}{$G_{RP}$} &
  \multicolumn{1}{c}{$\varpi$} &
  \multicolumn{1}{c}{$\sigma$} &
  \multicolumn{1}{c}{$D$} &
  \multicolumn{1}{c}{$\sigma$} &
  \multicolumn{1}{c}{Source} &
  \multicolumn{1}{c}{Notes}
  \\

  \multicolumn{1}{c}{} &
  \multicolumn{1}{c}{} &
  \multicolumn{1}{c}{deg} &
  \multicolumn{1}{c}{deg} &
  \multicolumn{1}{c}{days} &
  \multicolumn{1}{c}{} &
  \multicolumn{1}{c}{mag} &
  \multicolumn{1}{c}{mag} &
  \multicolumn{1}{c}{mag} &
  \multicolumn{1}{c}{mas} &
  \multicolumn{1}{c}{mas} &
  \multicolumn{1}{c}{kpc} &
   \multicolumn{1}{c}{kpc} &
  \multicolumn{1}{c}{} &
  \multicolumn{1}{c}{} 
  \\
  \hline
    AA Gem                      & 3430067092837622272 & 06:06:34.95 & +26:19:45.2 & 11.31285 & DCEP\_F & 9.363 & 9.956  & 8.62   & 0.2749 & 0.0177 & 3.81 & 0.11 & ESPaDOnS  & SOS,SOS \\
  AD Pup                      & 5614312705966204288 & 07:48:03.85 & -25:34:40.0 & 13.59738 & DCEP\_F & 9.541 & 10.125 & 8.786  & 0.2331 & 0.0165 & 0.82 & 0.04 & ESPaDOnS  & P21,DR3 \\
  AP Sgr                      & 4066429066901946368 & 18:13:02.50 & -23:07:02.3 & 5.05808  & DCEP\_F & 6.922 & 7.46   & 6.219  & 1.1815 & 0.024  & 2.85 & 0.12 & ESPaDOnS  & SOS,SOS \\
  AQ Pup                      & 5597379741549105280 & 07:58:22.09 & -29:07:48.3 & 30.15959 & DCEP\_F & 8.127 & 8.974  & 7.25   & 0.2751 & 0.0226 & 3.98 & 0.14 & ESPaDOnS  & SOS,SOS \\
  ASAS-SN J065046.50-085808.7 & 3050543819559526912 & 06:50:46.50 & -08:58:08.8 & 6.94564  & DCEP\_F & 13.49 & 14.402 & 12.547 & 0.1198 & 0.0179 & 2.66 & 0.10 & UVES,P110 & SOS,SOS \\ 

\hline\end{tabular}
\tablefoot{Meaning of the columns: Star = literature name of the DCEP. Gaia\_source\_id= $Gaia$ DR3 identifier. RA, Dec = equatorial co-ordinates at J2000. Period and Mode = period and mode of pulsation; DCEP\_F, DCEP\_1O, and DCEP\_MULTI refer to fundamental, first overtone, and multimode DCEPs, respectively. $G$, $G_{BP}$, and $G_{RP}$ are the magnitudes in the $Gaia$ bands. $\varpi$ and $\sigma$ are the $Gaia$ parallax and its error. $D$ and $\sigma$ are the distances and its error. Source shows the instrument used to observe the star. Notes report the origin of the periods (and modes) as well as of the photometry: ‘SOS’ means that the periods and the photometry come from $Gaia$ DR3 using the specific pipeline for the DCEP variables \citep[see][]{Ripepi2023}; ‘DR3’ means that the photometry is from the general DR3 source \citep[see][]{GaiaVallenari}, while ‘P21’ means that the periods are from      \citet{Piet2021}. Note that the distances for the four stars most likely classified as type II Cepheids are probably wrong (see Appendix~\ref{sect:typeII}). 
A portion of the Table is shown here for guidance regarding its form and content. The machine-readable version of the full table will be published at the Centre de Données astronomiques de Strasbourg (CDS, https://cds.u-strasbg.fr/).}
\end{sidewaystable}

In recent years, a still unresolved debate has taken place regarding the well-known discrepancy between the values of $H_0$ obtained by the SH0ES\footnote{Supernovae, HO, for the Equation of State of Dark energy} project through the cosmic distance ladder \citep[$H_0=$73.01$\pm$0.99 km s$^{-1}$ Mpc$^{-1}$,][]{Riess2022a} and the value computed by the Planck cosmic microwave background (CMB) project based on the flat $\Lambda$ cold dark matter ($\Lambda$CDM) model \citep[$H_0=$67.4$\pm$0.5 km s$^{-1}$ Mpc$^{-1}$,][]{Planck2020}. If no solution is found for this intricate $H_0$ tension, this could hint at the need for a revision of the $\Lambda$CDM model. It is therefore critical to evaluate the discrepancy through the analysis of the residual systematics \citep[see e.g.][and reference therein]{Dainotti2021,Freedman2021,Riess2021b}. 

Metallicity may play an important role as one of the possible residual sources of uncertainty in the cosmic distance ladder, influencing the calibration of the DCEP $PL$ relations. Since the shape and width of the DCEP instability strip are predicted to be affected by metallicity variations \citep[e.g.][]{Caputo2000}, a direct consequence should be seen when estimating the $PL$ relations coefficients \citep[][and references therein]{Marconi2005,Marconi2010,DeSomma2022}. To avoid systematic effects in the calibration of the extragalactic distance scale \citep[e.g.][]{Romaniello2008,Bono2010,Riess2016}, one has to also consider the metallicity dependence of the $PL$ relations and the reddening-free Wesenheit magnitudes \citep{Madore1982} in those cases in which the metallicity effect might be small, such as when dealing with the near-infrared regime \citep[NIR, see e.g.][]{Fiorentino2013,Gieren2018}.

In the past, the lack of accurate independent distances hampered direct empirical evaluations of the metallicity dependence of $PL$ relations using Galactic DCEPs with precise [Fe/H] measurements based on high-resolution (HiRes hereafter) spectroscopy, but this scenario completely changed with the advent of the \gaia\ mission \citep[][]{Gaia2016}. 
Accurate parallaxes were provided with data release 2 \citep[DR2, ][]{Gaia2018} and further improved with the early data release 3 \citep[EDR3, ][]{Gaia2021}, in addition to the discovery of hundreds of new Galactic DCEPs \citep[][]{Clementini2019,Ripepi2019,Ripepi2022a}. A substantial sample can thus be considered when other surveys are taken into consideration, such as the Optical Gravitational Lensing Experiment (OGLE) Galactic Disk survey \citep[][]{Udalski2018} and the Zwicky Transient Facility \citep[ZTF, ][]{Chen2020}, making it possible to improve not only studies related to the cosmic distance scale but also Galactic studies \citep[e.g.][and references therein]{Lemasle2022, Trentin2023a}.
Another constraint was the limited availability of HiRes measurements confined to the solar neighbourhood. This restriction resulted in a narrow range of [Fe/H] values centred on solar or slightly supersolar values, with a small dispersion of 0.2-0.3 dex \cite[e.g.][]{Genovali2014, Luck2018, Groenewegen2018, Ripepi2019}. Consequently, this limitation hindered the ability to generate results with statistical significance. 

To overcome these problems, we started a project named C-MetaLL\footnote{https://sites.google.com/inaf.it/c-metall/home} \citep[Cepheid - Metallicity in the Leavitt Law, see][for a full description]{Ripepi2021a}, with the primary goal of measuring the chemical abundance of a sample of at least 300 Galactic DCEPs through HiRes spectroscopy, as well as providing homogeneous multi-band time-series photometry. The main aim was to enlarge the iron abundance range towards the metal-poor regime -- that is, [Fe/H]$<-$0.4 dex -- in which only a few Galactic stars had abundance measurements in the literature. In the first two papers of the series \citep[][]{Ripepi2021a, Trentin2023a} (R21 and T23 from now on, respectively), we published accurate abundances for more than 25 chemical species for a total of 114 DCEPs, while the first results for the photometric part were presented in \citet{Bhardwaj2024} . In particular, in T23, we obtained measurements for 43 objects with [Fe/H]$<-$0.4 dex, reaching abundances as low as $-$1.1 dex. 
This paper represents the sixth manuscript in the series and a direct follow-up of T23, presenting the spectroscopic analysis of 180 new stars based on a total of 331 HiRes spectra and the study of the galactic gradient for 29 chemical elements.

While the main scope of this work is to study the metallicity dependence of the DCEP $PL$
relations in the context of the extragalactic distance scale, the sample of DCEPs presented in this work, in conjunction with our previous results, nevertheless amounts to a total of 294 DCEPs with a homogeneous derivation of stellar parameters and abundances. In addition, as we shall see later in this article, our sample is evenly distributed in metallicity, spanning a range from +0.5 dex to $-$1 dex in [Fe/H]  and 5 to 20 kpc in Galactocentric radius. The aforementioned characteristics make our sample an appropriate testing ground for the study of the Galactic disc abundance gradient, in comparison with literature works that in general use heterogeneous spectroscopic samples or datasets heavily unbalanced towards the solar vicinity \citep{Luck2018, lemasle2018milky, Minniti2020, Ripepi2022a, daSilva2023, Matsunaga2023}.   

The paper is organised as follows: in Sect. \ref{sec:observations}, we describe the sample of DCEPs and their properties; in Sect. \ref{sec:spec}, we describe the analysis technique; in Sect. \ref{sec:results} and Sect. \ref{sec:discussion}, we describe and discuss our results; and in Sect. \ref{sec:conclusion}, we outline our conclusions.

\section{Observations}\label{sec:observations}

\subsection{Sample}
\label{sect:sample}
The sample of DCEPs presented in this work was selected from the DCEP catalogue published in $Gaia$ data release 3 \citep[DR3][]{Ripepi2023}. We chose the pulsators with the intention of maximising the number of objects with low metallicity and long periods. This is because the main scope of the C-MetaLL survey is to derive accurate $PL$ and $PW$ relations for Galactic DCEPs. To this aim, we have had to enlarge the range covered by the independent variables in the quoted relations as much as possible; namely, periods and metallicities. Therefore, we have had to observe both short- and long-period DCEPs as well as metal-rich and metal-poor pulsators. While the periods are known, the metallicity can only be guessed based on the position of the stars in the disc, which, as was mentioned in the previous section, is well known to exhibit a metallicity gradient from the centre to the anticentre of the Galaxy. Thus, our observations aimed to maximise the most distant objects in the anticentre direction, possibly those with longer periods, which are rarer (since their evolutionary times are faster than those of shorter-period DCEPs). The list of the 180 targets is shown in Table~\ref{tab:programStars}, in which we list the main characteristics of the stars, including their periods, modes of pulsation, and distances. 
These last quantities were calculated based on the $PWZ$ relation published by \citet{Ripepi2022a} with
the procedure outlined in detail in Sect. 2.2 of \citet{GaiaDrimmel2023}. In brief, we used $Gaia$ photometry (see 
Table~\ref{tab:programStars}) to calculate the apparent Wesenheit magnitude $W=G-1.90\times(G_{BP}-G_{RP}$) \citep[][]{Ripepi2019}. Then, we inserted the periods and [Fe/H] values (measured in this paper) in the quoted $PWZ$ to obtain the absolute $W$ magnitude. The distance calculation was then straightforward, using the definition of distance modulus. The conversion in Galactocentric radii $R_{GC}$ was obtained in the usual way, adopting a distance of the Sun from the Galactic centre, $R_\odot = 8277 \pm 9$(stat) $\pm 30$(sys) pc \citep{GRAVITY2021} (for full details, we refer the reader to \citet{GaiaDrimmel2023}).

\begin{figure}
	\includegraphics[width=9cm]{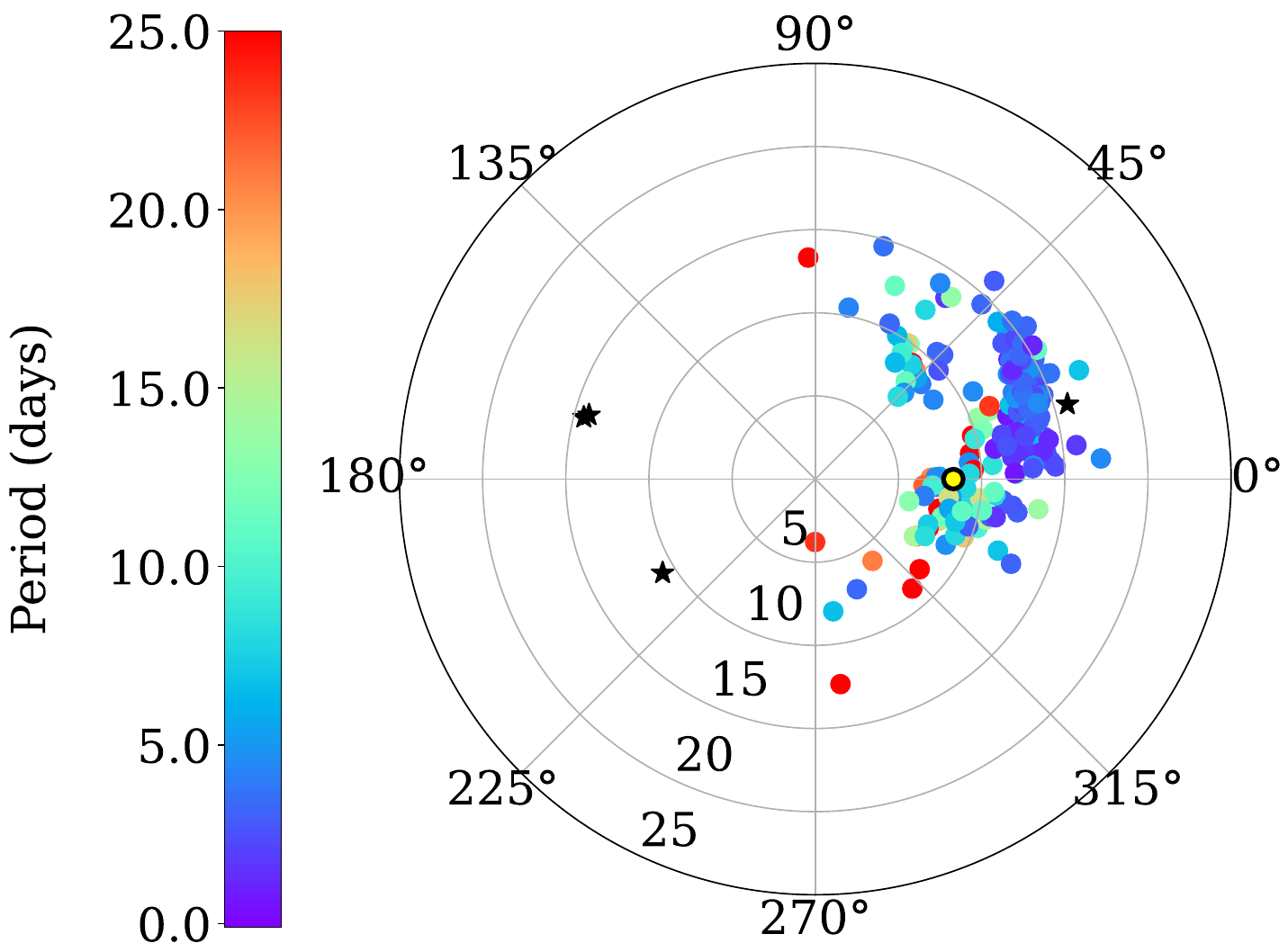}
    \caption{Map along the Galactic disc of the programme stars. The points are colour-coded according to the pulsator's period. The position of the sun is shown with a yellow-black circle. The four stars probably belonging to the type II Cepheid class are shown with black stars (see text).}
    \label{fig:mapTargets}
\end{figure}

The distribution of the targets along the Milky Way (MW) disc is displayed in Fig.~\ref{fig:mapTargets}. The distribution of the programme stars ranges from about 5 to 18 kpc in Galactocentric radii ($R_{GC}$) and covers all four Galactic quadrants, with a significant concentration in the third. 
In the figure, the stars are colour-coded according to their periods. The period distribution shows short-period DCEPs at larger Galactic radii, especially in the third quadrant. This is not a coincidence, but because there exists a period-age relation for DCEPs \citep[see e.g.][]{Bono2005} so that shorter periods indicate larger ages. It has been found that at larger Galactocentric radii the DCEPs are on average older \citep[e.g.][]{Skowron2019, desomma2020b, Desomma2021}. This explains the lack of long-period DCEPs at large radii in our sample. 
An additional feature displayed in Fig.~\ref{fig:mapTargets} is the presence of three stars with polar angles between 135$^o$ and 225$^o$; namely, Gaia DR2 4087335043492541696, V532 Sco, and ASAS-SN J165340.10-332041.7. Given the peculiar position along the Galactic disc, these pulsators may be type II Cepheids. An in-depth analysis of these objects is provided in Appendix~\ref{sect:typeII}. In addition, a fourth star, OGLE\,CEP-GD-0069, which was originally classified as DCEP in \citet{Udalski2018}, turned out to be a BL Her variable according to \citet{Ripepi2023}. This star is also discussed in the appendix. Given their uncertain nature, these four stars have been excluded from the following analysis.   

\begin{table*}
\footnotesize\setlength{\tabcolsep}{5pt}
\caption{Log of the observations for the 331 spectra analysed in this work.}
\label{tab:logObservations}
\begin{tabular}{lrrrrrrrrrl}
\hline
  \multicolumn{1}{c}{Star} &
  \multicolumn{1}{c}{HJD} &
  \multicolumn{1}{c}{Phase} &
  \multicolumn{1}{c}{Texp} &
  \multicolumn{1}{c}{S/N} &
  \multicolumn{1}{c}{$T_{eff}$} &
  \multicolumn{1}{c}{$\log g$} &
  \multicolumn{1}{c}{$\xi$} &
  \multicolumn{1}{c}{$V_{broad}$} &
  \multicolumn{1}{c}{$RV$} &
  \multicolumn{1}{c}{Source} \\

  \multicolumn{1}{c}{} &
  \multicolumn{1}{c}{days} &
  \multicolumn{1}{c}{} &
  \multicolumn{1}{c}{s} &
  \multicolumn{1}{c}{} &
  \multicolumn{1}{c}{K} &
  \multicolumn{1}{c}{dex} &
  \multicolumn{1}{c}{km {s$^{-1}$}} &
  \multicolumn{1}{c}{km {s$^{-1}$}} &
  \multicolumn{1}{c}{km {s$^{-1}$}} &
  \multicolumn{1}{c}{} \\

\hline
  AA Gem & 59269.76384 & 0.920 & 750 & 102 & 5656 $\pm$ 68 & 0.92 $\pm$ 0.05 & 3.0 $\pm$ 0.2 & 11.0 $\pm$ 1.0 & 0.9 $\pm$ 0.3 & ESPaDOnS\\
  AA Gem & 59630.79632 & 0.833 & 410 & 92 & 5877 $\pm$ 66 & 1.22 $\pm$ 0.05 & 3.3 $\pm$ 0.2 & 12.0 $\pm$ 1.0 & 0.8 $\pm$ 0.2 & ESPaDOnS\\
  AA Gem & 59838.14990 & 0.162 & 500 & 80 & 4887 $\pm$ 68 & 0.1 $\pm$ 0.05 & 2.8 $\pm$ 0.2 & 9.0 $\pm$ 1.0 & 24.6 $\pm$ 0.2 & ESPaDOnS\\
  AD Pup & 59269.81690 & 0.369 & 750 & 78 & 5128 $\pm$ 100 & 0.38 $\pm$ 0.05 & 2.7 $\pm$ 0.2 & 6.0 $\pm$ 1.0 & 74.0 $\pm$ 0.3 & ESPaDOnS\\
  AD Pup & 59630.82225 & 0.918 & 410 & 77 & 5983 $\pm$ 83 & 0.58 $\pm$ 0.05 & 3.4 $\pm$ 0.3 & 10.0 $\pm$ 1.0 & 68.3 $\pm$ 0.2 & ESPaDOnS\\
\hline\end{tabular}
\tablefoot{The different columns report: the name of the star, the heliocentric Julian day on which the spectrum has been collected, the phase, exposure time, S/N per pixel, effective temperature, logarithm of gravity, microturbulent velocity, broadening velocity, heliocentric radial velocity, and the instrument used to collect the HiRes spectroscopy. 
The phases were calculated adopting periods and epochs of maximum light from the $Gaia$ DR3 catalogue \citep[][]{Ripepi2023}, except for seven stars for which we report details on the phase determination in Appendix~\ref{sec:appPhase}.
A portion is shown here for guidance regarding its form and content. The machine-readable version of the full table will be published at the Centre de Données astronomiques de Strasbourg (CDS, https://cds.u-strasbg.fr/).}
\end{table*}

\begin{sidewaystable}[ht]
    \footnotesize\setlength{\tabcolsep}{5pt}
    \caption{Estimated chemical abundances for the 180 analysed DCEPs.}
    \label{tab:abundances}
    \begin{tabular}{crrrrrrrrrrl}
    \hline
        \multicolumn{1}{c}{Star}&
        \multicolumn{1}{c}{[C/H]} & 
        \multicolumn{1}{c}{[O/H]} & 
        \multicolumn{1}{c}{[Na/H]} & 
        \multicolumn{1}{c}{[Mg/H]} & 
        \multicolumn{1}{c}{[Al/H]} & 
        \multicolumn{1}{c}{[Si/H]} & 
        \multicolumn{1}{c}{[S/H]} & 
        \multicolumn{1}{c}{[Ca/H]} & 
        \multicolumn{1}{c}{[Sc/H]} & 
        \multicolumn{1}{c}{[Ti/H]} &
        \multicolumn{1}{c}{...}\\
    \hline
    AA Gem & $-$0.54 $\pm$ 0.03 & $-$0.04 $\pm$ 0.09 & 0.16 $\pm$ 0.1 & 0.14 $\pm$ 0.05 & 0.11 $\pm$ 0.09 & $-$0.24 $\pm$ 0.1 & $-$0.12 $\pm$ 0.02 & $-$0.09 $\pm$ 0.02 & 0.23 $\pm$ 0.06 & 0.04 $\pm$ 0.08 & ... \\
    AD Pup & $-$0.41 $\pm$ 0.11 & 0.16 $\pm$ 0.11 & 0.16 $\pm$ 0.19 & 0.39 $\pm$ 0.03 & 0.14 $\pm$ 0.11 & $-$0.22 $\pm$ 0.11 & $-$0.06 $\pm$ 0.01 & $-$0.19 $\pm$ 0.09 & 0.17 $\pm$ 0.09 & 0.05 $\pm$ 0.11 & ... \\
    AP Sgr  & $-$0.22 $\pm$ 0.11 & 0.2 $\pm$ 0.11 & $-$0.03 $\pm$ 0.16 & 0.1 $\pm$ 0.1 & 0.23 $\pm$ 0.11 & 0.01 $\pm$ 0.1 & 0.17 $\pm$ 0.02 & 0.05 $\pm$ 0.05 & 0.3 $\pm$ 0.06 & 0.09 $\pm$ 0.1 & ... \\\\

    \hline
        \multicolumn{1}{c}{...}&
        \multicolumn{1}{c}{[V/H]} & 
        \multicolumn{1}{c}{[Cr/H]} & 
        \multicolumn{1}{c}{[Mn/H]} & 
        \multicolumn{1}{c}{[Fe/H]} & 
        \multicolumn{1}{c}{[Co/H]} & 
        \multicolumn{1}{c}{[Ni/H]} & 
        \multicolumn{1}{c}{[Cu/H]} & 
        \multicolumn{1}{c}{[Zn/H]} & 
        \multicolumn{1}{c}{[Sr/H]} & 
        \multicolumn{1}{c}{[Y/H]} & 
        \multicolumn{1}{c}{...}\\
    \hline
    ...& 0.05 $\pm$ 0.07 & $-$0.09 $\pm$ 0.04 & $-$0.28 $\pm$ 0.04 & $-$0.15 $\pm$ 0.07 & $-$0.17 $\pm$ 0.05 & $-$0.02 $\pm$ 0.09 & 0.03	$\pm$ 0.1 & 0.04 $\pm$ 0.2 & 0.21 $\pm$ 0.09 & $-$0.29 $\pm$ 0.08 & ... \\
    ... & 0.06 $\pm$ 0.07 & $-$0.13 $\pm$ 0.08 & $-$0.32 $\pm$ 0.1 & $-$0.13 $\pm$ 0.08 & 0.06 $\pm$ 0.06 & $-$0.13 $\pm$ 0.11 & $-$0.2 $\pm$ 0.11 & $-$0.35 $\pm$ 0.11 & 0.28 $\pm$ 0.11 & $-$0.22 $\pm$ 0.09 & ... \\
    ... & 0.0 $\pm$ 0.1 & $-$0.02 $\pm$ 0.09 & $-$0.26 $\pm$ 0.05 & $-$0.02 $\pm$ 0.09 & 0.62 $\pm$ 0.04 & 0.25 $\pm$ 0.11 & $-$0.15 $\pm$ 0.13 & $-$0.18 $\pm$ 0.13 & 0.25 $\pm$ 0.11 & $-$0.19 $\pm$ 0.01 & ... \\\\

    \hline
        \multicolumn{1}{c}{...}& 
        \multicolumn{1}{c}{[Zr/H]} &
        \multicolumn{1}{c}{[Ba/H]} & 
        \multicolumn{1}{c}{[La/H]} & 
        \multicolumn{1}{c}{[Ce/H]} & 
        \multicolumn{1}{c}{[Pr/H]} & 
        \multicolumn{1}{c}{[Nd/H]} & 
        \multicolumn{1}{c}{[Sm/H]} & 
        \multicolumn{1}{c}{[Eu/H]} & 
        \multicolumn{1}{c}{[Gd/H]} &
        \multicolumn{1}{c}{Source}\\
    \hline
    ... & 0.24 $\pm$ 0.34 & 0.04 $\pm$ 0.05 & 0.22	$\pm$ 0.07 & 0.16 $\pm$	0.09 & $-$0.3 $\pm$ 0.09 & 0.15	$\pm$ 0.07 & 0.12 $\pm$0.05 & ---~~~~~~~ & 0.22	$\pm$ 0.09& ESPaDOnS \\
    ... & 0.17 $\pm$ 0.41	& 0.32 $\pm$ 0.11 & 0.02 $\pm$ 0.11	& $-$0.01 $\pm$ 0.11	& $-$0.4 $\pm$ 0.11	& 0.15 $\pm$ 0.09 & 0.14 $\pm$ 0.1 & ---~~~~~~~ & 0.31 $\pm$ 0.11 & ESPaDOnS \\
    ... & 0.32 $\pm$ 0.37	& 0.21 $\pm$ 0.11 & $-$0.02 $\pm$ 0.02 & $-$0.13 $\pm$ 0.11 & $-$0.21 $\pm$ 0.11 & 0.09 $\pm$ 0.1 & 0.09 $\pm$ 0.09	& ---~~~~~~~ & $-$0.03 $\pm$ 0.11& ESPaDOnS \\\\
    \hline\end{tabular}
    \tablefoot{The table is divided into three parts. The first and last columns report the name of the star and the source, respectively (see Table \ref{tab:logObservations} and \ref{tab:programStars}). The other columns report the estimated abundances (and relative errors) in solar terms for the chemical species analysed in this work. The machine-readable version of the full table will be published at the Centre de Données astronomiques de Strasbourg (CDS, https://cds.u-strasbg.fr/).}
\end{sidewaystable}

\subsection{Instruments used for the observations and data reduction}\label{Sec:instruments}

For the observations, three instruments were used: 

\begin{itemize} 

\item  The Ultraviolet and Visual Echelle Spectrograph \citep[UVES,][]{UVES_Dekker2000}\footnote{\ https://www.eso.org/sci/facilities/paranal/instruments/uves.html}), attached at Unit Telescope 2 (UT2) of Very Large Telescope (VLT), placed at Paranal (Chile). The red arm was used, equipped with the grism CD\#3, covering the wavelength interval 4760--6840 {\AA}, and with the central wavelength at 5800 {\AA}. The 1 arcsec slit, which provides a dispersion of R$\sim$47,000 (sampling 2~px), was selected for all the targets. 

\item The High Accuracy Radial velocity Planet Searcher for the Northern hemisphere \citep[HARPS-N,][]{HARPS_Mayor2003,HARPS_Cosentino2012} \footnote{https://www.tng.iac.es/instruments/harps/}), attached at the 3.5m Telescopio Nazionale Galileo (TNG). HARPS-N features an echelle spectrograph covering the wavelength range between 3830 to 6930 {\AA}, with a spectral resolution of R=115,000 (sampling 3.3~px). 

\item The Echelle SpectroPolarimetric Device for the Observation of Stars (ESPaDOnS\footnote{ https://www.cfht.hawaii.edu/Instruments/Spectroscopy/Espadons/}) attached at the 3.6m Canada-France-Hawaii Telescope (CFHT). ESPaDOnS provides a spectral resolution of R=81,000 (sampling 0.69~px) in the wavelength range between 3700 and 10500 {\AA}.

\end{itemize}

Typically, we obtained 1-2 epoch spectra with UVES, 2-3 with CFHT, and three or more with HARP-N. The signal-to-noise ratio (S/N) was larger than 40 for about 80\% of the spectra. In total, in this work, we analysed 331 spectra for 180 DCEPs. A complete list of the time of acquisition and the individual S/N values is reported in Table~\ref{tab:logObservations}. 

Reduction of all the spectra, which included bias subtraction, spectrum extraction, flat-fielding, and wavelength calibration, was done by automatic pipelines provided by the three instrument teams so that we downloaded the science-ready one-dimensional spectra \citep[for more details on the data reduction of HARPS-N, UVES and ESPaDOnS spectra, see][respectively]{Ripepi2021a,Trentin2023a,Bhardwaj2023}.

Finally, from Table~\ref{tab:programStars} it can be noted that the stars OGLE-GD-CEP-0026, OGLE-GD-CEP-1278, and OGLE-GD-CEP-1290 have been observed with both the UVES and HARP-N instruments. These stars are therefore useful for cross-checking the results obtained with the different instruments (see Sect. \ref{sec:results}).

\begin{figure*}
	\includegraphics[width=\textwidth]{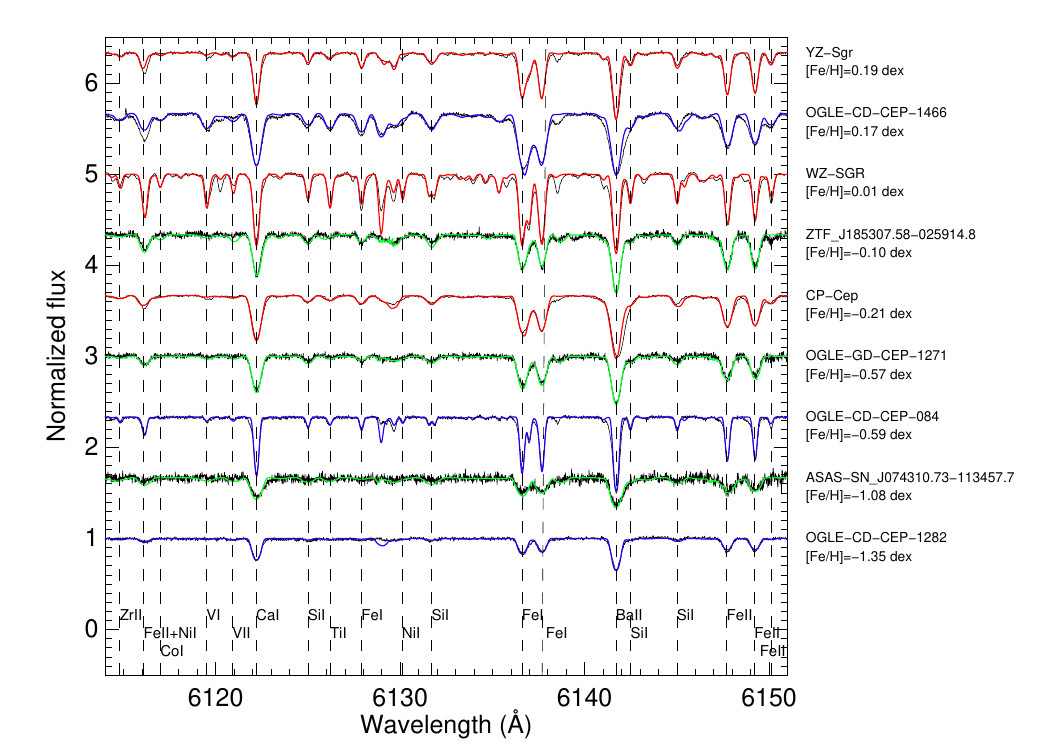}
    \caption{Example of spectral synthesis for our targets in the region from $\lambda$\, = 6110 {\AA} to $\lambda$\, = 6151 {\AA}. The colours differ according to the particular instrumental equipment, specifically in red the ESPaDOnS spectra, in green the HARPS spectra, and in blue the UVES spectra. The spectra are ordered, from top to bottom, by decreasing metallicity. The main spectral lines have been identified at the bottom.}
    \label{fig:spectra}
\end{figure*}

\begin{figure}
    \centering
    \includegraphics[width=9cm]{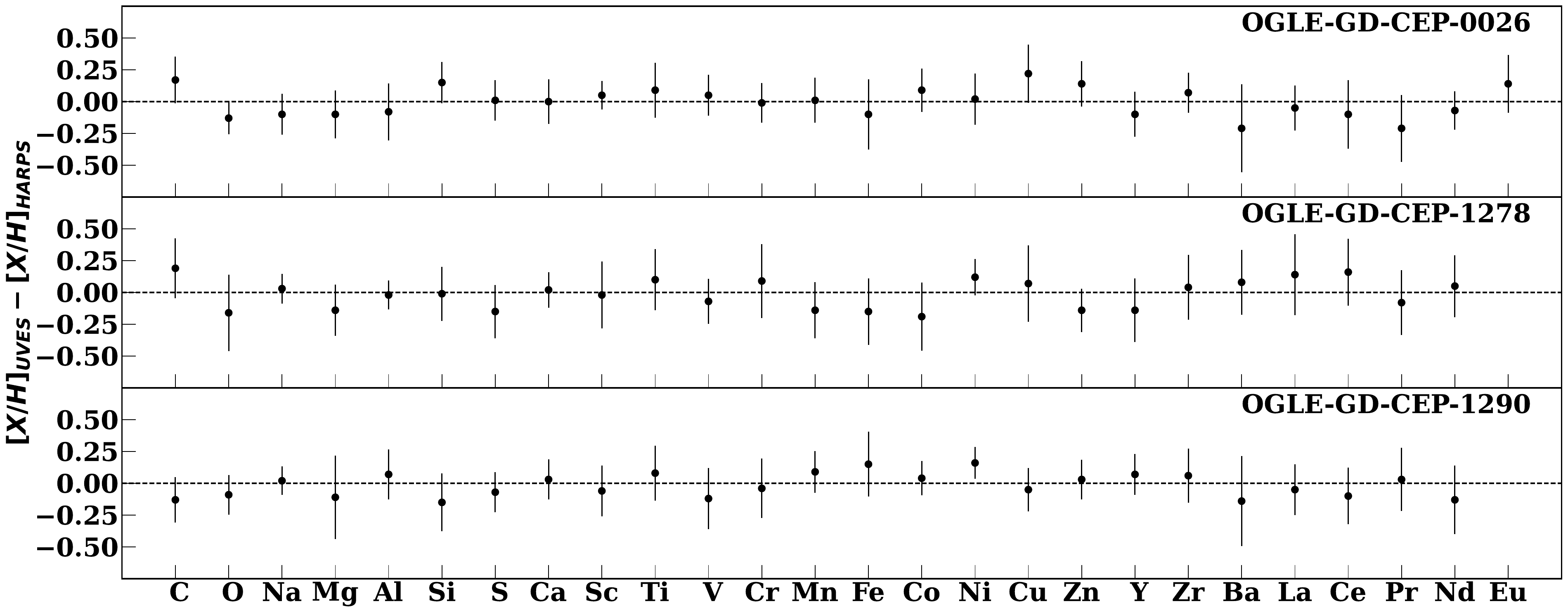}
    \caption{Differences between the UVES and HARPS chemical abundances for each star observed with both instruments.}
    \label{fig:elem_comp}
\end{figure}

\begin{figure*}
\sidecaption
	\includegraphics[width=12cm]{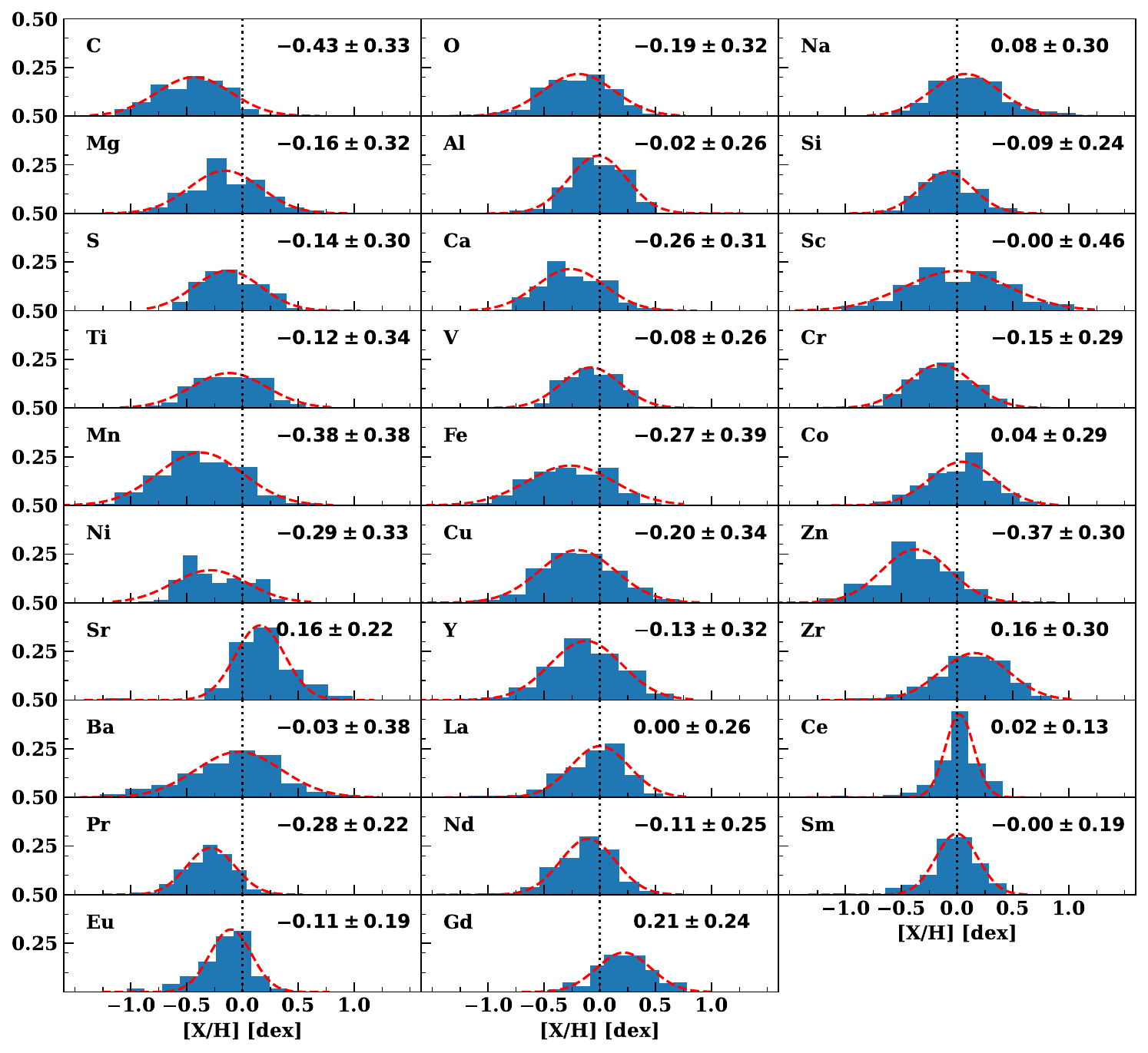}
    \caption{Histograms of the distribution of the chemical abundances derived in this study. Gaussian fits have been superimposed (dashed red lines) with the respective mean value and FWHM reported in each panel.}
    \label{fig:hist}
\end{figure*}

\begin{figure*}
\sidecaption
	\includegraphics[width=12cm]{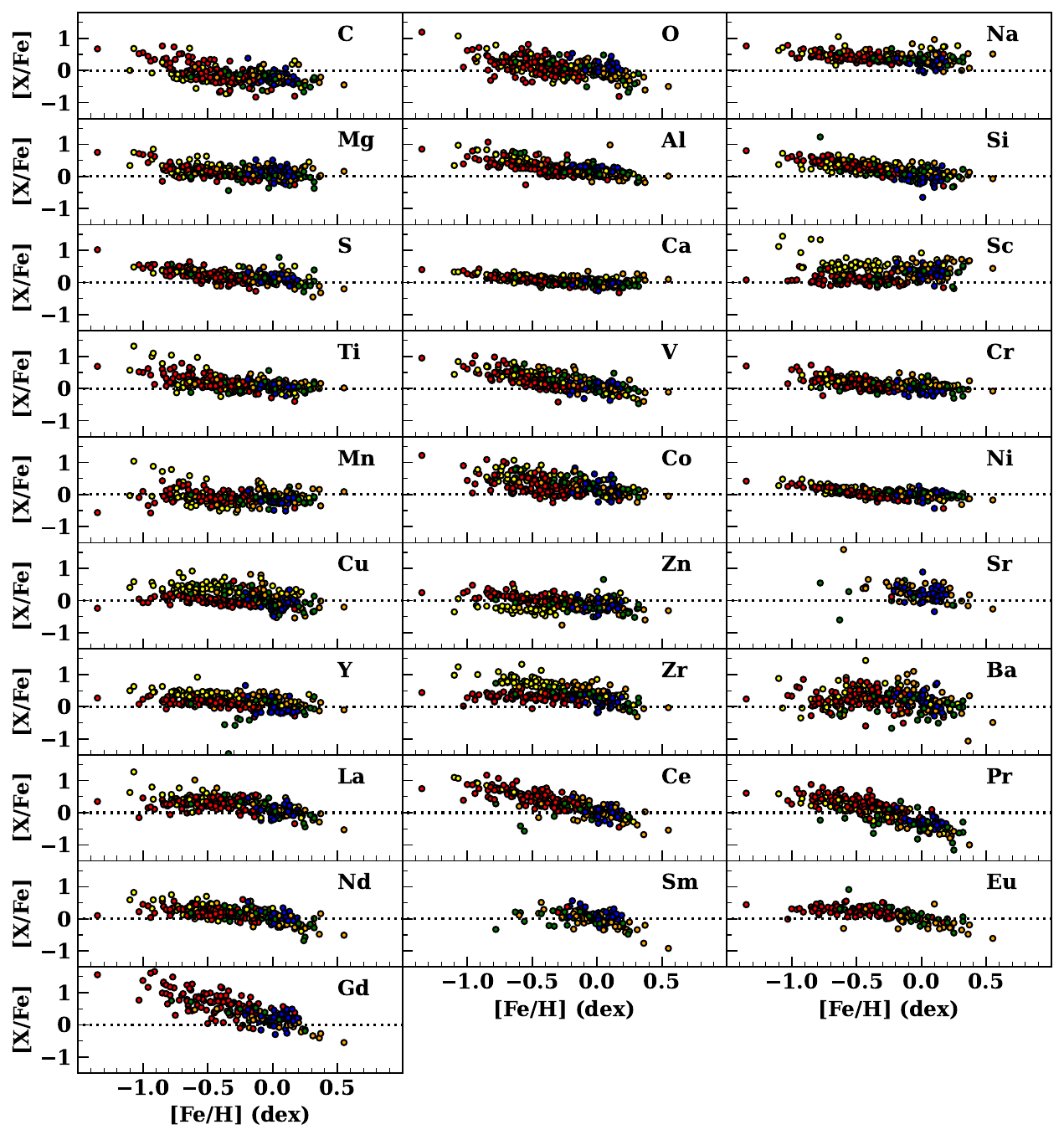}
    \caption{Chemical elements in the form of [X/Fe] plotted against the iron content. The colours differ according to the particular instrumental equipment, specifically in red the UVES spectra, in green the HARPS spectra, and in blue the ESPaDOnS spectra. The targets from R21 and T23 are depicted in orange and yellow, respectively (see also Fig.~\ref{fig:alpha_vs_iron} and \ref{fig:iron_grad} for the colour legend).}
    \label{fig:elem_vs_iron}
\end{figure*}

\begin{figure}
	\includegraphics[width=9cm]{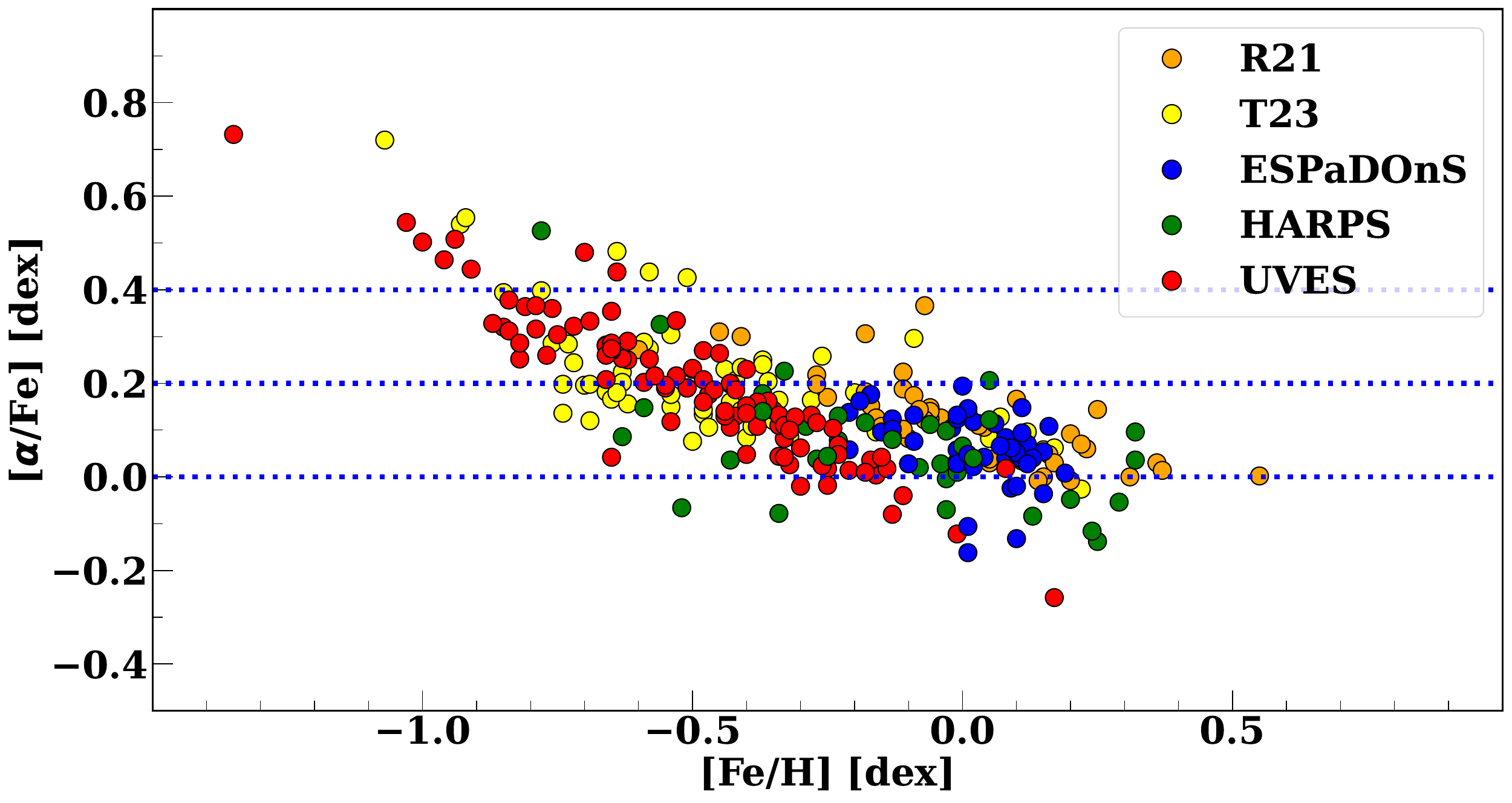}
    \caption{$\alpha$-elements abundance in the form of [$\alpha$/Fe] plotted against the iron content. The colour pattern, the same as in Fig.~\ref{fig:elem_vs_iron}, is described in the upper right legend. The horizontal dotted blue lines highlight [$\alpha$/Fe] abundances at 0.4, 0.2, and 0 dex. Typical errors for the abundances are around 0.15 dex.}
    \label{fig:alpha_vs_iron}
\end{figure}

\section{Spectroscopy}\label{sec:spec}
\subsection{Stellar parameters}
The adopted method for our spectroscopic analysis is the same as was used in T23; here, we only report the main procedure.
The initial step in measuring chemical abundances involves determining the primary atmospheric parameters, including the effective temperature (T${\rm eff}$), surface gravity ($\log g$), microturbulent velocity ($\xi$), and the line-broadening parameter ($V_{broad}$), defined as the full width at half maximum (FWHM) of a Gaussian line-broadening function representing the combination of macroturbulence and rotational velocity (with macroturbulence often playing a dominant role in the case of DCEPs).

A tool widely adopted in the literature for estimating effective temperature is the line depth ratio (LDR) method \citep{kovtyukh2000precise,gray1991precise}. This method offers the advantage of being sensitive to temperature variations, while remaining unaffected by abundance changes and interstellar reddening. Typically, for each spectrum in our targets, we measured approximately 32 LDRs listed in \citet{kovtyukh2000precise}. The mean values and weighted standard deviation were used as the final temperatures and errors.

For the remaining parameters ($\xi$ and $\log g$), we employed an iterative approach. The estimation of microturbulence involved ensuring that iron abundances showed no dependence on equivalent widths (EWs); that is, that the slope of [Fe/H] against EWs is null. To achieve this occurrence, we initially measured the EWs of 145 \ion{Fe}{I} lines using a custom {\tt Python} semi-automatic routine. The line sample was extracted from the list published by \citet{romaniello2008influence}, and the routine minimised errors in continuum estimation on the spectral line wings. EWs were converted to abundances using the WIDTH9 code \citep{kurucz1981solar}, applied to the corresponding atmospheric model calculated using ATLAS9 \citep{kurucz1993new}. In this calculation, the influence of $\log g$ was not considered, as neutral iron lines are insensitive to it.
Subsequently, surface gravities were estimated through a similar iterative procedure, imposing ionisation equilibrium between \ion{Fe}{I} and \ion{Fe}{II}. The adopted list of 24 \ion{Fe}{II} lines was extracted from \citet{romaniello2008influence}. Errors were estimated through the propagation of errors derived from the linear fits.

\subsection{Abundances}
To circumvent issues arising from spectral line blending due to line-broadening, a spectral synthesis technique was applied to our spectra. Synthetic spectra were generated in three steps: i) plane-parallel local thermodynamic equilibrium (LTE) atmosphere models were computed using the ATLAS9 code \citep{kurucz1993new}, employing the stellar parameters; ii) stellar spectra were synthesised using SYNTHE \citep{kurucz1981solar}; and iii) the synthetic spectra were convoluted to account for instrumental and line-broadening. This convolution was evaluated by matching the synthetic line profiles to a selected set of observed metal lines.

For a total of 29 different chemical elements including Fe, it was possible to detect the spectral lines used for the estimation of the abundances. For all elements, we performed the following analysis: we divided the observed spectra into intervals, 25 Å or 50 Å wide, and derived the abundances in each interval by performing a $\chi^2$ minimisation of the differences between the observed and synthetic spectra. The minimisation algorithm was written in {\tt Python}, using the {\tt amoeba} routine. 

We considered several sources of uncertainties in our abundances. First, we evaluated the expected errors caused by variations in the fundamental stellar parameters of $\delta T_{\mbox{\scriptsize eff}} = \pm 150$ K, $\delta \log g= \pm$\,0.2~dex, and $\delta\xi = \pm$\,0.3~km\,s$^{-1}$. 
According to our simulations, those errors contribute $\approx \pm$\,0.1~dex to the total error budget. Total errors were evaluated by summing in quadrature the value obtained by the error propagation and the standard deviations obtained from the average abundances. For those elements for which only one spectral line could be detected, we evaluated the error only by considering error propagation on the fundamental stellar parameters.

The adopted lists of spectral lines and atomic parameters were taken from \citet{castelli2004spectroscopic}, who updated the original parameters of \citet{kurucz1995kurucz}. When necessary, we also checked the NIST database \citep{ralchenko2019nist}. In Fig.~\ref{fig:spectra}, we plot an example of some of the spectra analysed in this work. More details about the spectral lines that can be typically detected and used to estimate chemical abundances can be found in R21 and T23. The final list of LTE abundances for all the stars is reported in Table~\ref{tab:abundances}. All the abundances are referred to the solar value (\citet{grevesse2011chemical}).

\begin{figure}
	\includegraphics[width=9cm]{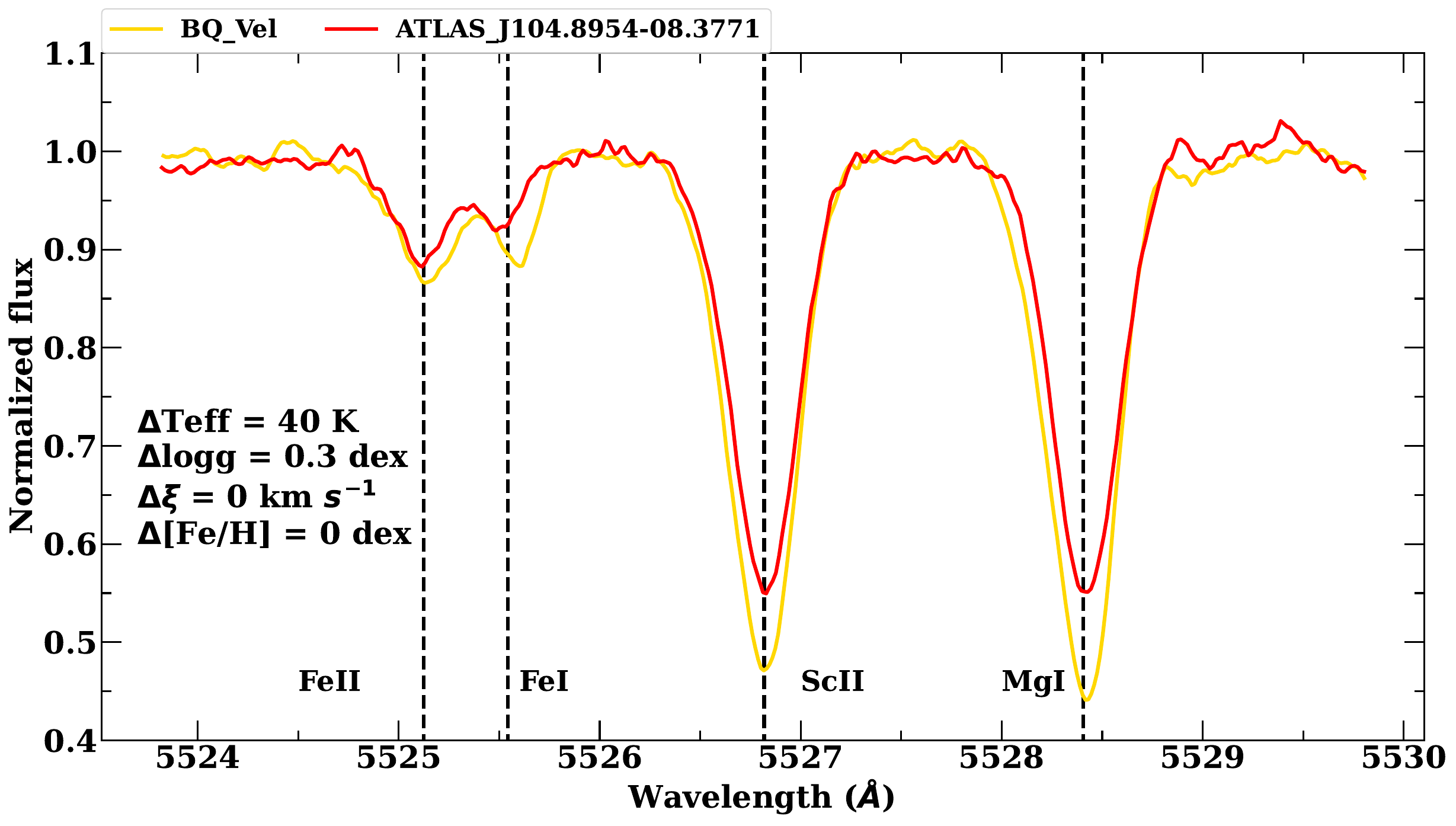}
    \caption{Example of two spectra around the \ion{Sc}{II} line at $\lambda\lambda$ 5526.818$\AA$ for the stars BQ Vel (in yellow from the T23) and ATLAS-J104.8954-08.3771 (in red from the UVES sample analysed in this paper). Other spectral lines are highlighted with dashed black lines. Atmospheric parameters and metallicity differences between the two targets are reported.}
    \label{fig:comparison_sc}
\end{figure}

\begin{figure}
	\includegraphics[width=9cm]{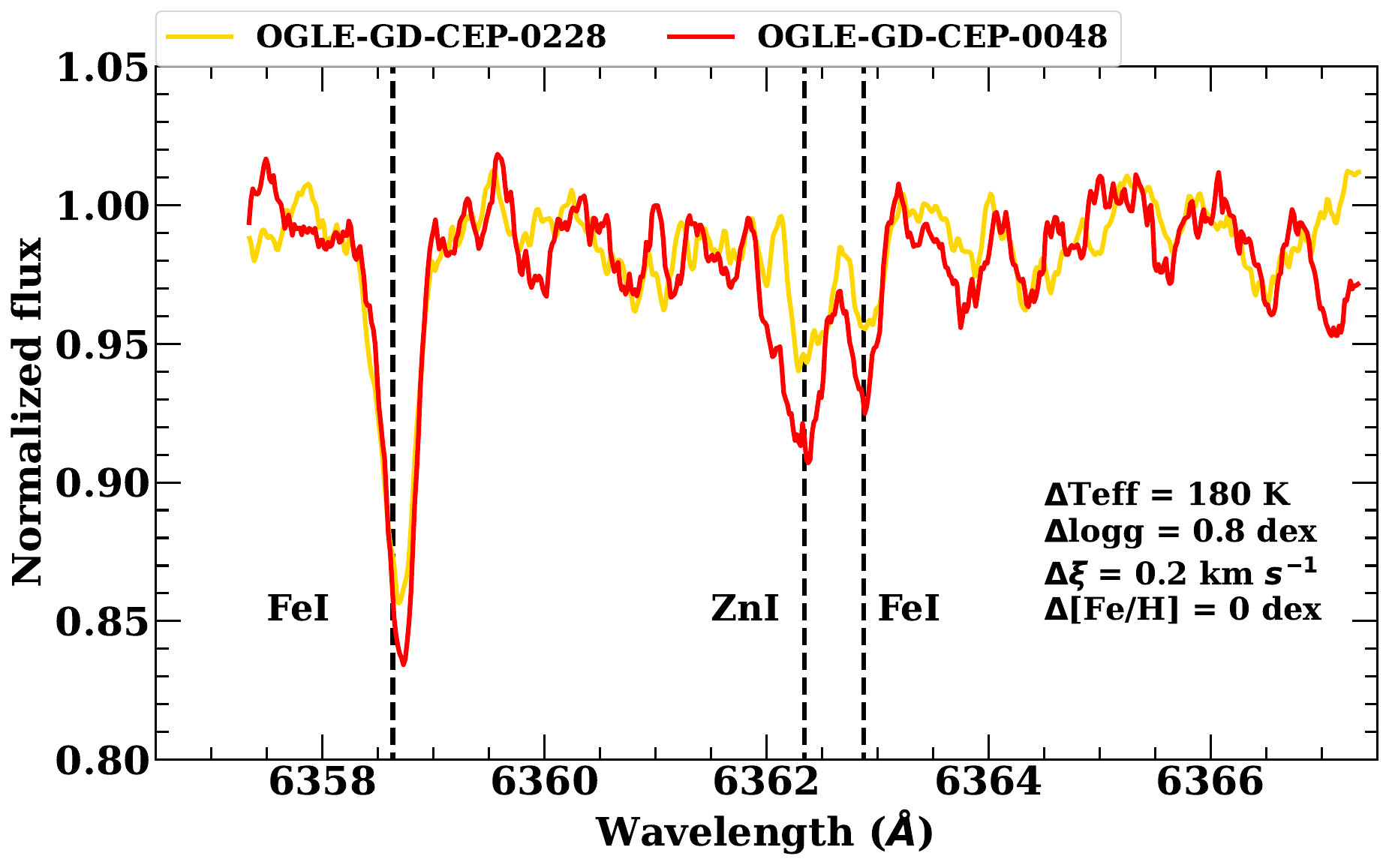}
    \caption{Example of two spectra around the \ion{Zn}{I} line at $\lambda\lambda$ 6362.338$\AA$ for the stars OGLE-GD-CEP-0228 (in yellow from the T23) and OGLE-GD-CEP-0048 (in red from the UVES sample analysed in this paper). Other spectral lines are highlighted with dashed black lines. Atmospheric parameters and metallicity differences between the two targets are reported.}
    \label{fig:comparison_zn}
\end{figure}

\section{Results}\label{sec:results}
\subsection{Abundances results}
As a first step in the analysis of our results, we verified that for the three targets observed with both UVES and HARPS-N and mentioned in Sect. \ref{Sec:instruments} we find consistent results among the estimated abundances. In this perspective, the homogeneity in the analysis method should not be influenced significantly by the use of the different instruments to observe our targets. As is shown in Fig.~\ref{fig:elem_comp}, for all three stars we find comparable chemical values between the UVES and HARPS spectra and no systematics are detected.

In Fig.~\ref{fig:hist}, we show histogram distributions of each derived element abundance, for a total of 29 different chemical species. Bins have been fixed to 0.15 dex, reflecting typical error values. Each distribution has been fitted with a Gaussian curve. It is worth noting that most of these Gaussian curves have FWHM > 0.25 dex, which means that for these elements we have an almost homogeneous distribution over a broad range of values. 
Moreover, when confronting current Fig.~\ref{fig:hist} with Fig. 3 of T23, it is possible to appreciate how this new sample better covers the whole range of abundances without sharp ‘jumps’ among bins that appeared in T23 for elements (see for example the  O, Ca, and Fe panels).
We complemented our sample with those presented in the previous C-MetaLL papers, R21 and T23, reaching a total of 292 pulsators analysed homogeneously. 

In Fig.~\ref{fig:elem_vs_iron}, we plot each element (in its [X/Fe] form) in relation to iron. Generally, we confirm all the trends already found in R21 and T23, in which we divided the chemical species into groups according to their main channel of formation within the stellar evolution models. In more detail, for light and $\alpha$ elements (from C to Ti), we observe a descending behaviour at lower metallicities until [Fe/H]\,$\approx$\,-0.5 dex,  with a flattening effect at higher values. Moreover, we note that while in T23 the region with over solar abundances at lower [Fe/H] was poorly populated with stars that appeared as outliers (yellow points), in this work we can confirm a negative trend that flattens at higher iron abundances. In Fig.~\ref{fig:alpha_vs_iron}, we have plotted the $\alpha$ elements versus [Fe/H] (to estimate $\alpha$, we used the average between [Mg/Fe], [Si/Fe], [S/Fe], [Ca/Fe], and [Ti/Fe]), confirming the $\alpha$ enhancement at lower metallicities that had already been observed in the literature from both an observational  \citep{Hayden2015,duffau2017,Trentin2023a} and a theoretical \citep{tinsley1979,matteucci1990,palicio2023} context. 
It is important to highlight that [Fe/H] and [$\alpha$/Fe] depends on both age and birth radius \citep[see for example ][]{wielen1996birth,ness2019galactic}. Since all the objects are Cepheids, the relation illustrates the [Fe/H] dependence of [$\alpha$/Fe] at a fixed age. In other words, [Fe/H] represents in this case the (birth) Galactocentric radius rather than age.

We note that for the Sc element, most of the low-metallicity stars studied in this paper appear systematically less abundant (around solar values) and with a flatter distribution than those presented in T23, which are clearly over-abundant and with a slight descending trend towards higher [Fe/H] values. In order to investigate this discrepancy between the two samples, we selected stars with similar atmospheric parameters (and possibly similar [Fe/H] abundances) and compared the spectra around the available scandium lines. An example with the two targets, BQ Vel from T23 and ATLAS-J104.8954-08.3771, is shown in Fig.~\ref{fig:comparison_sc}. It was indeed found that stars from T23 show more intense Sc lines (that is, higher abundances) than those presented in this work. This suggests that this effect could be real and not caused by systematics in the analysis. Possible effects due to different instruments are also excluded since most of the stars involved were observed with UVES.

Iron peak elements (from V to Zn) present a descending trend over the range of metallicities, with the exceptions of Mn and Zn, which appear flatter. It is worth noting how for the latter element the objects presented in this work, observed with UVES (red points), populate the over-solar and solar abundance zone at lower metallicities, while most of the previous targets from T23 (also observed with UVES) were found to be under-solar. This is a similar but opposite behaviour to the Sc case. In the wavelength range covered with the UVES instrument, however, there is only one available neutral zinc line at $\lambda\lambda$ 6362.338 Å. Although this line is weak and could be difficult to detect when the S/N is not optimal, a difference in the line depth has indeed been observed as in the case of Sc (see Fig.~\ref{fig:comparison_zn}). A similar study has been performed for the Copper, for which there are two neutral lines, at $\lambda\lambda$ 5105.5 and 5218.201 Å. In this case, the UVES targets are slightly under-abundant with respect to the other samples, but again a visual inspection confirmed this difference.

For heavier elements (from Sr to Gd), we either find a descending behaviour (Y, Ce, Pr, Gd) or a sort of arch-form trend with a bump at [Fe/H]\,$\approx$\,$-$0.2 dex ( Ba, La, Nd, and Eu), after which we find that abundances decrease with increasing metallicities. This curved trend was already found in the literature related to open cluster (OC) studies \citep[e.g.][]{molero2023} and is typical of those elements produced via the s-process. While the descending branch at higher metallicities is caused by the iron production from SNe Ia, at lower metallicities pollution from low- and intermediate-mass stars during the AGB phase cause the rising trend. 
    Given the low number of stars, we cannot draw a firm conclusion for both Sr and Sm. As happened for Sc, Cu, and Zn, and also for the zirconium (\ion{Zr}{II} line at $\lambda\lambda$ 6114.853Å), there seems to be a sort of bifurcation at the low-metallicity tail. We point out that we found the arch-form trend for some of the r-process elements as well (i.e. La and Eu).

\begin{figure}
	\includegraphics[width=9cm]{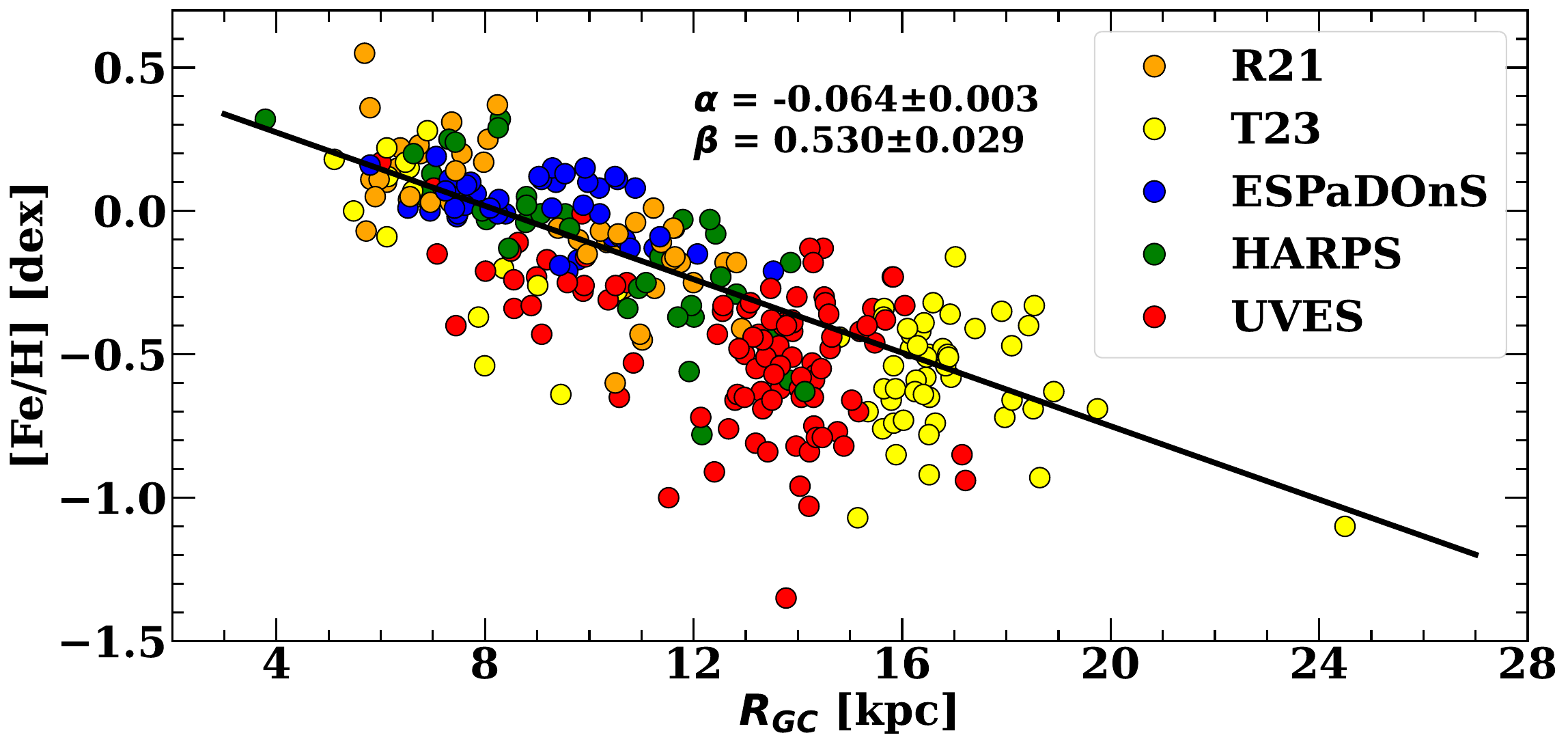}
    \caption{Galactic iron radial gradient. The colours are the same as in Fig.~\ref{fig:elem_vs_iron}. The black line represents the linear fit computed. Results of the fit are reported both in the figure and in Table~\ref{tab:coeff}. Typical errors for [Fe/H] are around 0.15 dex.}
    \label{fig:iron_grad}
\end{figure}

\begin{figure}
	\includegraphics[width=9cm]{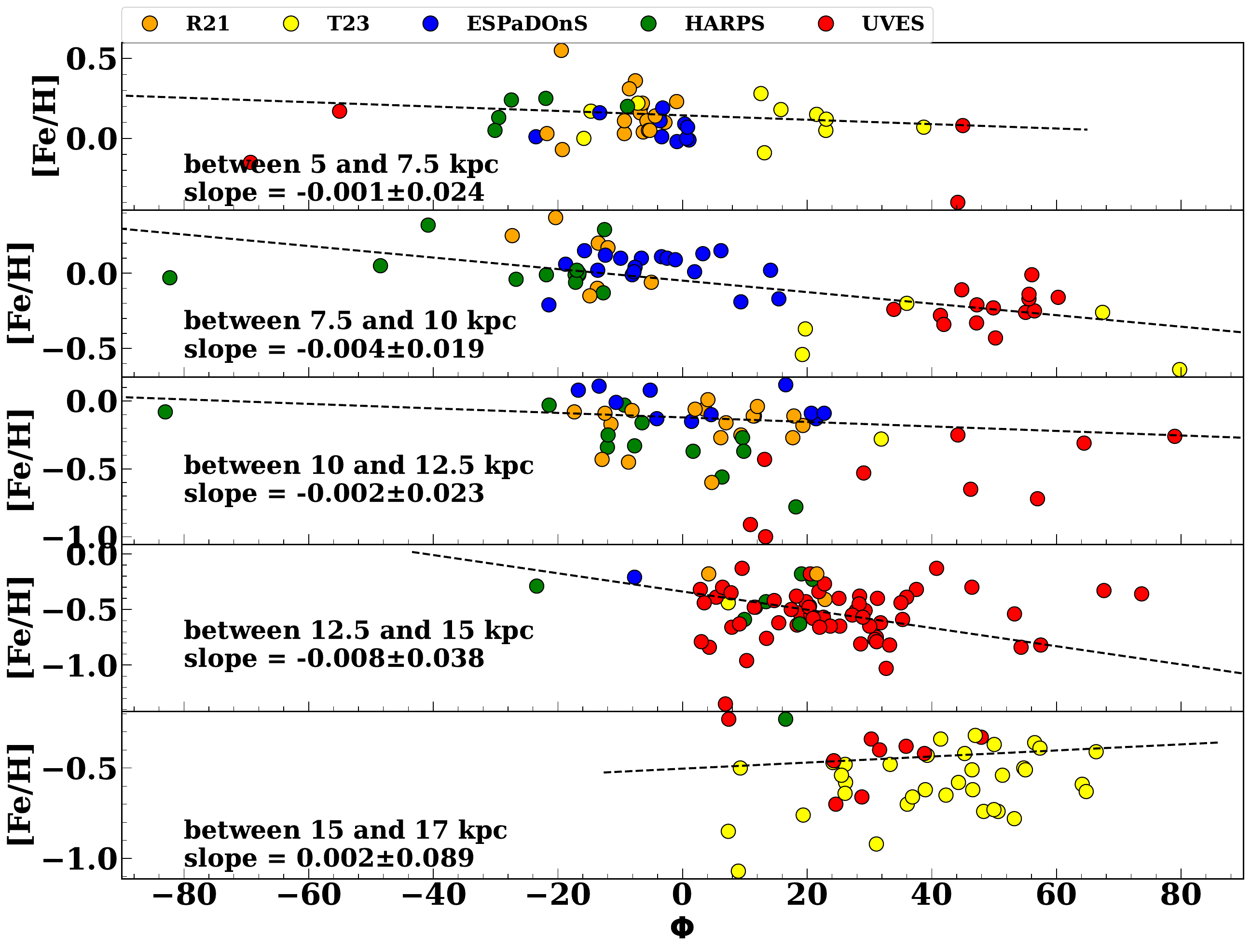}
    \caption{Galactic iron azimuthal gradient for different $R_{GC}$ ranges. The colours are the same as in Fig.~\ref{fig:elem_vs_iron}. The dashed black line represents the linear fit computed. In each panel, results of the slope are reported as well as the $R_{GC}$ range.}
    \label{fig:phi_grad}
\end{figure}

\begin{figure*}
\sidecaption
	\includegraphics[width=12cm]{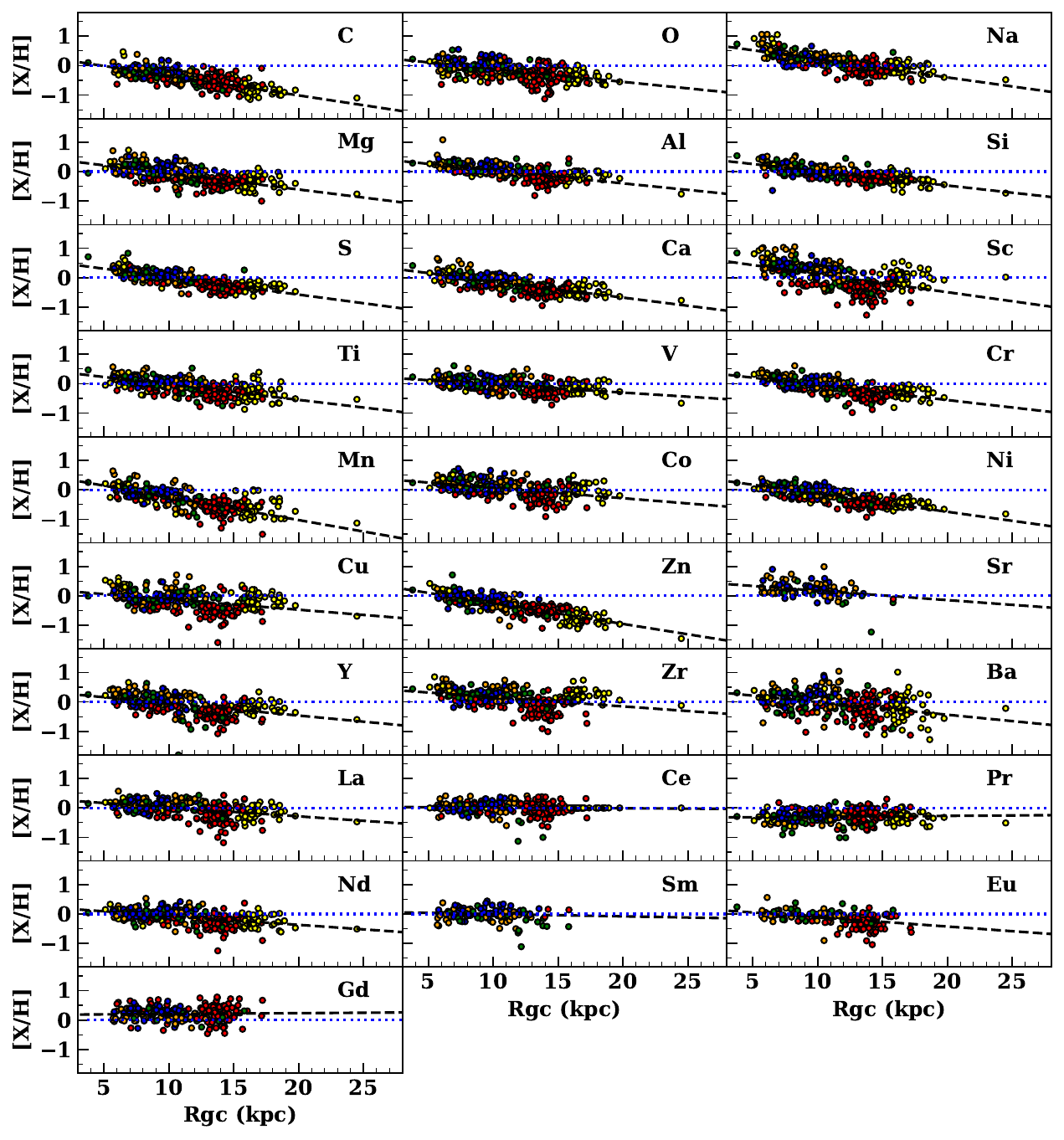}
    \caption{Galactic radial gradient for all the estimated chemical elements. The lines and colours are the same as in Fig.~\ref{fig:iron_grad}. The linear coefficients are listed in Table~\ref{tab:coeff} }
    \label{fig:elem_grad}
\end{figure*}

\subsection{Radial gradients and spatial distribution}

 Following the same procedure described in R21 and T23 and outlined in Sect.~\ref{sect:sample}, we estimated the Galactocentric distances, $R_{GC}$, and estimated the radial gradients for all the available chemical species.
All of the fits were carried out with the {\tt python LtsFit} package \citep{Cappellari2013}, which allows one to use weights on both axes and implements a robust outlier removal and error estimation of the fitted parameters.
In Fig.~\ref{fig:iron_grad}, we plot the iron radial Galactic gradient. As was expected, we found a clear negative trend. Moreover, an almost homogeneous distribution in metallicity is reflected in a homogeneous distribution in terms of distance, with the stars spanning a Galactocentric distance between 5 and 20 kpc.  Since in T23 we lacked a fair amount of stars with distances between 12 and 16 kpc and we had to complement the sample with literature stars, we might find different results between the current work and the previous one. One first difference is indeed the lack of a visible break at around 9.25 kpc. This is confirmed when trying to compute a two-line fit that overlaps almost perfectly with the  current result. From a quantitative point of view, we did not note a big discrepancy in the value of the slope. In more detail, we found a slightly higher value (in an absolute sense) but still comparable with both the single-line case and the inner slope of the two-line fit. Similarly, when considering the results in T23 obtained after a binning division of the sample, to compensate for the non-uniform distribution, the overall slope is in agreement with our most recent study (there is a good agreement with the two-line fit as well). This puts a further emphasis on the importance of having access to as uniform and homogeneous a sample of stars as possible in terms of distance (and metallicity), one of the key points of the C-MetaLL project. Similarly to what happened in T23, we found some outliers quite far below the linear fit. These objects were observed mostly with UVES (red and yellow points) but stars observed with HARPS (orange and green points) are found as well. Since all the spectra were analysed with the same method and the measurement errors do not justify their behavior, we can consider these outliers to be real.
In Fig.~\ref{fig:elem_grad}, we show the radiant galactic gradient for the other 28 elements. The estimated coefficients, together with the root mean square (rms), are listed in Table ~\ref{tab:coeff}. These results generally agree with those listed in T23. As was already seen in T23, Ba has the largest dispersion. Other elements (i.e. Sc, Ti, and Zn) present instead a difference of $>$0.01 dex kpc$^{-1}$. In these cases, we tried to repeat the linear regression including the literature sample used in T23 as well, obtaining results in perfect agreement with those of this work, once again highlighting the role of the farthest (and most metal-poor) objects and their weight in the linear fit. This is further highlighted when considering the significant difference in the number of objects used in the two works, counterbalanced by the homogeneity of the analysis and the sampling of the stars along the Galactocentric radii. 

\begin{figure}
	\includegraphics[width=9cm]{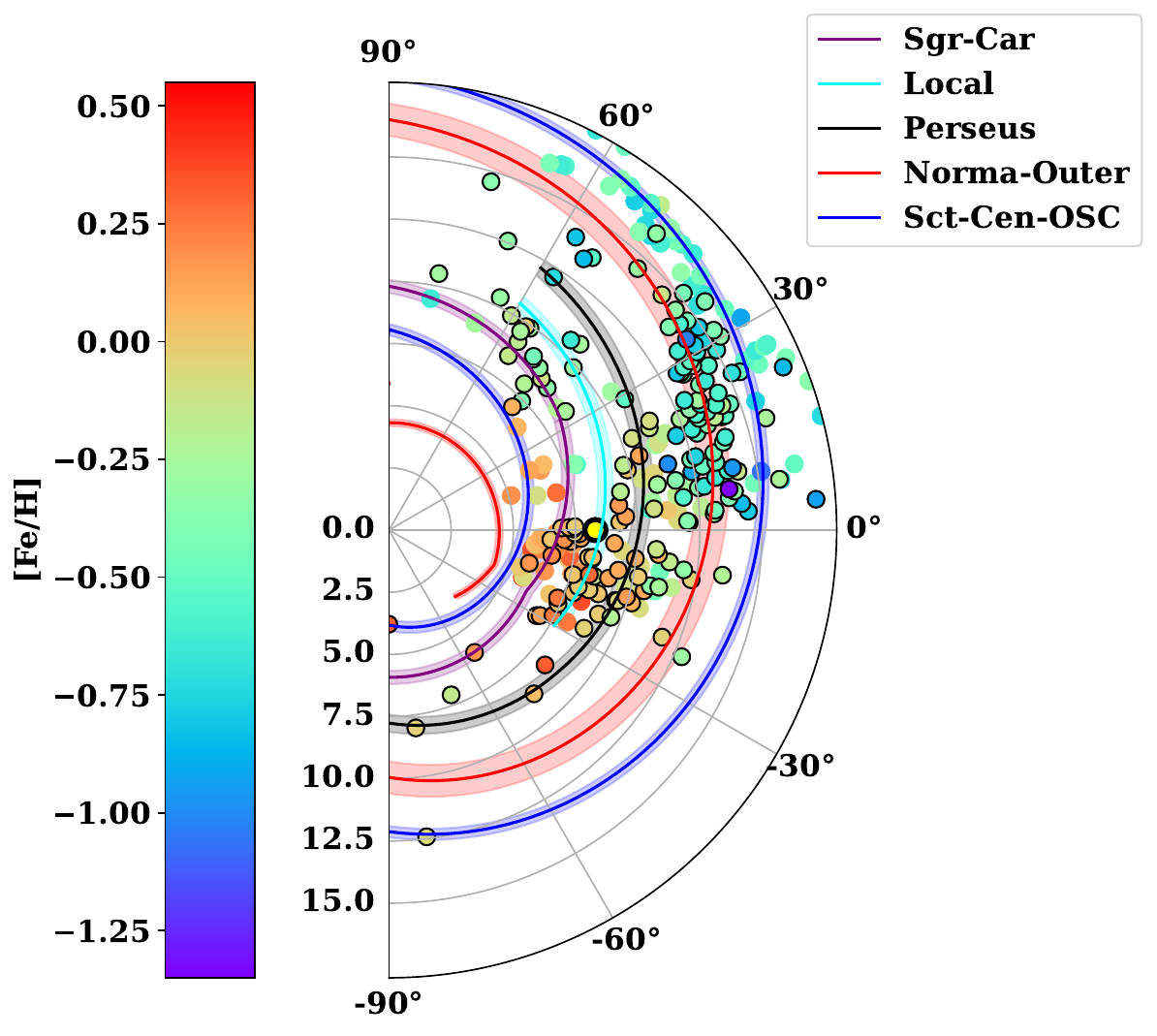}
    \caption{Galactic polar distribution of the stars. The new targets presented in this work are highlighted with black borders. The points are colour-coded according to [Fe/H] values and specified on the colour bar. The position of the sun is shown with a yellow-black symbol. Spiral arms from \citet{Reid2019} are superimposed. The colours are explained in the upper right legend.}
    \label{fig:spatial}
\end{figure}

To achieve more information about the distribution in the Galaxy of our sample, we have plotted our targets in polar coordinates in Fig.~\ref{fig:spatial}. We computed the azimuthal gradient at different $R_{GC}$ but we did not find significant results (see Fig.~\ref{fig:phi_grad}).

\begin{figure*}
\sidecaption
	\includegraphics[width=12cm]{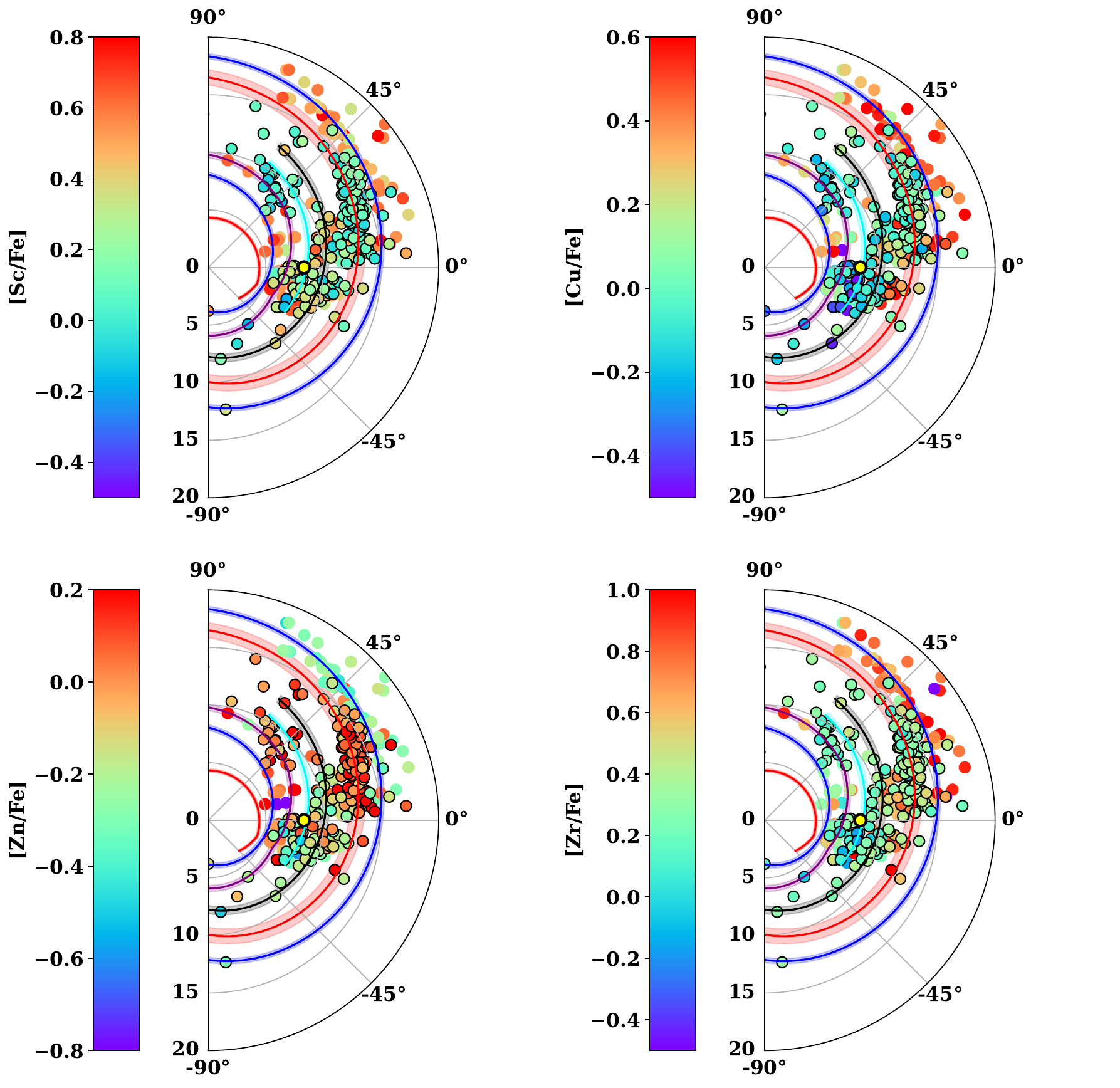}
    \caption{Galactic polar distribution of the stars. The new targets presented in this work are highlighted with black borders. The points are colour-coded according to the chemical element specified on the colour bar. The position of the sun is shown with a yellow-black symbol. We alert the reader to the different colour bar scales for each element.}
    \label{fig:polar_distr}
\end{figure*}

Given the peculiar distribution already discussed in T23, we superimposed the spiral arms described in \citet{Reid2019} (a similar study has been done by \citet{drimmel2024}). As was already done in \citet{minniti2021using} for stars beyond the Galactic bulge, we extended the outermost spiral structures and obtained a qualitative fit for the furthest stars. The slight adjustments to the spiral structure parameters are within the 1$\sigma$ errors reported in Table 2 by \citet{Reid2019}. In more detail, we notice that at distances larger than 12.5 kpc, most of the targets studied in this work seem to finely trace the Outer arm, while those presented in T23 seem to better fit the extension to the second and third Galactic quadrants of the Sct-Cen-OSC arm (and not the Outer one stated in T23). The fit presented in this paper could be a pivotal piece of information in extending the OSC arm, which is poorly constrained in this region of the MW \citep[see e.g.][]{Dame2011,sun2015,Reid2019,minniti2021using}.
Taking into consideration this information, we can find a possible justification for the anomalous behaviour of the DCEPs studied here concerning the Sc, Cu, Zn, and Zr abundances that have been discussed in the previous section. As is shown in Fig.~\ref{fig:polar_distr}, the evident and abrupt abundance change for the outermost stars could be because these two ‘blocks’ of stars are tracing two different spiral arms (Outer and OSC) and not only the Outer arm stated in T23.\footnote{ Note that the stars investigated in T23 were on average slightly more distant than the ones studied in this work} No firm conclusion could be drawn about the opposite behaviour of Zn compared with the other three elements.
 Lastly, it is worth noting that minor differences may also be seen in the interval 6-11 kpc (for example in Sc, Cu, and Zr) between the different samples of our project. Since we can exclude a bias between the different samples (see e.g. Fig.~\ref{fig:comparison_sc} and ~\ref{fig:comparison_zn} and the relative discussion), these differences may be real and explained by the association of our stars with different spiral arms.


\section{Comparison with recent literature results}\label{sec:discussion}
In this section, we compare our results with those found in the literature in recent years. Similar to what was done in T23, we focus our attention on studies involving either Cepheids or OCs, since they span a wide range of ages and represent optimal tracers. Since our slope is in agreement with our previous work, we refer to this paper for a comparison with less recent literature works to avoid redundancies in the discussion. 

Recently, \citet{daSilva2023} collected 356 DCEPs and compared the linear and logarithmic regression along the Galactocentric distance axis for [Fe/H], [O/H], and [S/H], finding a change of slope that flattens after around 12.5 kpc, a feature not found in this work. Regarding the linear fit, their values strongly disagree with ours:\\
O : $(-0.029\pm 0.006) R_{GC} +(0.21\pm0.05)$\\
S: $(-0.081\pm0.004) R_{GC} +(0.59\pm0.03)$\\
Fe:$(-0.041\pm0.003) R_{GC} + (0.32\pm0.02)$\\
As was already stated in their work, a possible explanation could be related to the different models used to derive abundances (see their section 4.1) and distances (see their section 2.3). Another cause of this discrepancy could be related to the different amount of targets available at $R_{GC}$>15kpc. Nonetheless, similarly to what was stated in their work, we find a strong difference between the O and S gradients, the latter being considerably steeper than the former. 

In the near-infrared \textit{YJ} bands, \citet{Matsunaga2023} reveal for the first time a gradient through the analysis of 16 cepheids, with a possible steepening of the gradient for $R_{GC}$> 5.6 kpc. Although their fit for $R_{GC}$<15 kpc DCEPs ([Fe/H]= $(-0.050\pm0.003)R_{GC} + (0.423\pm0.029)$ including literature stars from \cite{Luck2018} results in slightly steeper slope than \cite{daSilva2023}, it is still in disagreement with our slope by more than 2$\sigma$. We point out how in this case the work of \citet{Matsunaga2023} is mostly focussed on the inner Galactic disc.

In a recent study related to OCs, \citet{Magrini2023} use a sample of objects available in the final release of the $Gaia-$ESO survey, for a total of 67 open clusters. Their slope from a single line fit, [Fe/H] = $(-0.054\pm0.004)R_{GC} + (0.474\pm0.045)$, is  flatter by more than 1$\sigma$. On the other side, when considering a two-line fit with a break at $R_{GC}$ = 11.2kpc, an agreement with their inner slope is found ([Fe/H] = $(-0.081\pm0.008)R_{GC} + (0.692\pm0.068)$) only at 2$\sigma$. 
Although their outer slope is considerably flatter ([Fe/H] =$(-0.044\pm0.014) R_{GC} + (0.376\pm0.178)$), there is still statistical agreement. 

A similar discussion applies to the study by \citet{yogesh2024}, who collected a remarkable sample of 1879 OCs to study the evolution of metallicity in the Galactic disc. About 90\% of their objects are younger than 1 Gyr and the resulting slope (for $R_{GC}$<12.8kpc) is in agreement with our work ($-0.070 \pm 0.002$ dex $kpc^{-1}$) at 1.5 $\sigma$.

In their investigation into the origin of neutron-capture elements, \citet{molero2023} derived chemical evolution models, starting with the two-infall model by \citet{chiappini1997}, and confronted them with observational data from the 6th Data Release of the $Gaia-$ESO survey. Their observation (with OCs) and prediction (Model R-150) for the present-time radial gradient for [Fe/H] and [Eu/H] are in good agreement with our results (see their table 2).  

Except for a slightly shallower slope documented in \citet{netopil2022} ($-0.058$ dex kpc$^{-1}$), derived from a merged dataset of individual studies and the Apache Point Observatory Galactic Evolution Experiment (APOGEE) survey comprising 136 OCs, our findings align well with several other OC studies previously explored in depth by \citet{Magrini2023}, specifically those by \citet{spina2022} and \citet{zhang2021a}, which report slopes of $-0.064\pm0.007$ and $-0.066\pm0.005$, respectively. 

On the contrary, \citet{myers2022} identify a somewhat steeper slope ($-0.073\pm0.002$ dex kpc$^{-1}$), although one that is statistically consistent when focussing solely on young OCs (age < 1Gyr) (refer to table 2 of \cite{Magrini2023}). Of note are the findings of \citet{Magrini2023}, particularly in their tables A.8 and A.10, for which linear weighted fits were conducted for various chemical species. Overall, strong concordance is observed both in the overall fit and when restricting the analysis to young OCs (age < 1Gyr).

\section{Summary}\label{sec:conclusion}

In the context of the C-MetaLL project, we have presented the results obtained from 331 HiRes spectra collected with the UVES/VLT, HARPS-N/TNG, and ESPaDOnS/CFHT instruments for a sample of 180 individual DCEPs located over a wide range of Galactocentric radii (5$<R_{GC}<20$ kpc). For each target, we derived accurate atmospheric parameters, radial velocities, and abundances for up to 29 different species. The iron abundances range between +0.5 and $-$1 dex with a rather even distribution of DCEPs with metallicity. 
The sample presented in this paper was complemented with the data already published in the context of the C-MetaLL survey, resulting in a total of 292 pulsators whose spectra have been analysed in a homogeneous way. 

In this work, we have exploited our homogeneous sample to study the abundance gradients of the Galactic disc in a range of Galactocentric radii spanning the range of 5-20 kpc. Here, we report the main results of this work:
\begin{itemize}
    \item Studying the relation between each chemical species (in its [X/Fe] form) and iron, we observe an $\alpha$ enhancement for [Fe/H]$<-$0.5 dex. This is a well-known occurrence, already observed both empirically \citep{Hayden2015,duffau2017,Trentin2023a} and theoretically \citep{tinsley1979,matteucci1990,palicio2023}. At higher metallicities, this negative trend flattens.
    \item Iron peak and heavy elements generally present a negative gradient compared with the iron content, while those produced via the s-process (namely, Zr, Ba, and La) show an arch-form behaviour \citep{molero2023}, with a change of trend at [Fe/H]\,$\approx$\,$-$0.2 dex.
    \item For some elements (Sc, Cu, Zn, and Zr), we have found an apparent bias in the abundances between T23 and this study at lower metallicities, with Zn having an opposite behaviour to what appears for Sc, Cu, and Zr. Based on a direct visual comparison of the spectra for stars with similar atmospheric parameters and iron content, we have confirmed the difference in the line depths for the available spectral lines for these elements. 
    \item As was expected, we have found a clear negative radial gradient for most of the elements (except Ce, Pr, Sm, and Gd). Our targets are evenly distributed over a broad range of distances. When complementing our sample with other literature stars from other works (as was done in T23, for example), the slopes do not significantly change. Good agreement is found with other DCEP and OC works in the most recent literature.
    \item Focussing on the case of iron, we did not find any evidence for a break at any galactic radii and the estimated slope ($-0.064 \pm 0.003$ dex kpc$^{-1}$) is in good agreement compared with both the single-line case of \citet{Trentin2023a} and the inner slope (when considering their two-line case). 
    \item Superimposing the polar distribution of our dataset with the Galactic spiral arms \citep{Reid2019,minniti2021using}, we found a qualitative fit for our farthest star, which traces both the Outer and the extension of OSC arms. Should this result be confirmed in further works, it would be possible to put better constraints on both the spiral structures in a region that is still poorly explored (in particular on the OSC arm). 
    \item The possible association of our targets with the two spiral arms mentioned above could qualitatively justify the apparent change of abundance trend for Sc, Cu, Zn, and Zr. 
    Interestingly enough, the polar representation of the distribution of these four elements shows a peculiar behaviour of zinc. This species appears more abundant in the stars associated with the Outer arm, while the opposite is visible for Sc, Cu, and Zr. No clear explanation has been found for this behaviour.  
\end{itemize}

Finally, the homogeneous sample of DCEPs with metallicities from HiRes spectroscopy will be exploited in a forthcoming paper for the study of the metallicity dependence of the DCEP $PL$ relations.

\begin{table}
    \centering
    \caption{Results of the fitting.} 
    \centering
    \begin{tabular}{lccc}
    \hline
    El & $\alpha$  & $\beta$ & rms \\ 
       & (dex kpc$^{-1}$) & (dex) & (dex)\\
    \hline
C & -0.067 $\pm$ 0.003 & 0.320 $\pm$ 0.035 & 0.13 \\ 
O & -0.044 $\pm$ 0.004 & 0.329 $\pm$ 0.050 & 0.19 \\ 
Na & -0.061 $\pm$ 0.003 & 0.828 $\pm$ 0.039 & 0.16 \\ 
Mg & -0.055 $\pm$ 0.004 & 0.488 $\pm$ 0.047 & 0.19 \\ 
Al & -0.044 $\pm$ 0.003 & 0.486 $\pm$ 0.035 & 0.12 \\ 
Si & -0.049 $\pm$ 0.002 & 0.501 $\pm$ 0.029 & 0.08 \\ 
S & -0.059 $\pm$ 0.003 & 0.596 $\pm$ 0.032 & 0.13 \\ 
Ca & -0.055 $\pm$ 0.003 & 0.434 $\pm$ 0.035 & 0.15 \\ 
Sc & -0.061 $\pm$ 0.006 & 0.736 $\pm$ 0.071 & 0.33 \\ 
Ti & -0.052 $\pm$ 0.004 & 0.488 $\pm$ 0.043 & 0.16 \\ 
V & -0.028 $\pm$ 0.003 & 0.259 $\pm$ 0.040 & 0.16 \\ 
Cr & -0.051 $\pm$ 0.003 & 0.458 $\pm$ 0.038 & 0.14 \\ 
Mn & -0.078 $\pm$ 0.004 & 0.529 $\pm$ 0.044 & 0.17 \\ 
Fe & -0.064 $\pm$ 0.003 & 0.530 $\pm$ 0.029 & 0.11 \\
Co & -0.035 $\pm$ 0.004 & 0.422 $\pm$ 0.053 & 0.20 \\ 
Ni & -0.061 $\pm$ 0.003 & 0.478 $\pm$ 0.034 & 0.11 \\ 
Cu & -0.035 $\pm$ 0.005 & 0.236 $\pm$ 0.059 & 0.26 \\ 
Zn & -0.070 $\pm$ 0.003 & 0.451 $\pm$ 0.035 & 0.11 \\ 
Sr & -0.032 $\pm$ 0.009 & 0.496 $\pm$ 0.088 & 0.15 \\ 
Y & -0.041 $\pm$ 0.004 & 0.366 $\pm$ 0.049 & 0.22  \\ 
Zr & -0.031 $\pm$ 0.005 & 0.474 $\pm$ 0.057 & 0.23 \\ 
Ba & -0.043 $\pm$ 0.006 & 0.429 $\pm$ 0.073 & 0.33 \\ 
La & -0.030 $\pm$ 0.004 & 0.319 $\pm$ 0.044 & 0.19 \\ 
Ce & -0.002 $\pm$ 0.002 & 0.037 $\pm$ 0.026 & 0.10 \\ 
Pr & 0.003 $\pm$ 0.004 & -0.329 $\pm$ 0.044 & 0.13 \\ 
Nd & -0.031 $\pm$ 0.003 & 0.245 $\pm$ 0.039 & 0.16 \\ 
Sm & -0.008 $\pm$ 0.008 & 0.077 $\pm$ 0.077 & 0.15 \\ 
Eu & -0.032 $\pm$ 0.005 & 0.209 $\pm$ 0.054 & 0.13 \\ 
Gd & 0.003 $\pm$ 0.006 & 0.178 $\pm$ 0.071 & 0.21 \\ 

    \hline
    \hline
    \end{tabular}
    \tablefoot{Coefficients of the linear fit of the form [X/H]=$\alpha\times R_{GC}+\beta$ with relative dispersion coefficient.}
    \label{tab:coeff}
\end{table}

\begin{acknowledgements}
This research has made use of the
SIMBAD database operated at CDS, Strasbourg, France.
We acknowledge funding from INAF GO-GTO grant 2023 “C-MetaLL - Cepheid metallicity in the Leavitt law” (P.I. V. Ripepi). 
 This work has made use of data from the European Space
Agency (ESA) mission Gaia (https://www.cosmos.esa.int/gaia),
processed by the Gaia Data Processing and Analysis Consortium (DPAC,
https://www.cosmos.esa.int/web/gaia/dpac/consortium). Funding
for the DPAC has been provided by national institutions, in particular, the
institutions participating in the Gaia Multilateral Agreement.
This research was supported by the Munich Institute for Astro-, Particle and BioPhysics (MIAPbP), which is funded by the Deutsche Forschungsgemeinschaft (DFG, German Research Foundation) under Germany´s Excellence Strategy – EXC-2094 – 390783311.
This research was supported by the International Space Science Institute (ISSI) in Bern/Beijing through ISSI/ISSI-BJ International Team project ID \#24-603 - “EXPANDING Universe” (EXploiting Precision AstroNomical Distance INdicators in the Gaia Universe).
\end{acknowledgements}

\bibliographystyle{aa} 
\bibliography{myBib} 
-------------------------------------------------------------

\begin{appendix} 

\section{Stars with uncertain classification}
\label{sect:typeII}

\begin{figure*}
\centering
\hbox{
\includegraphics[width=6.0cm]{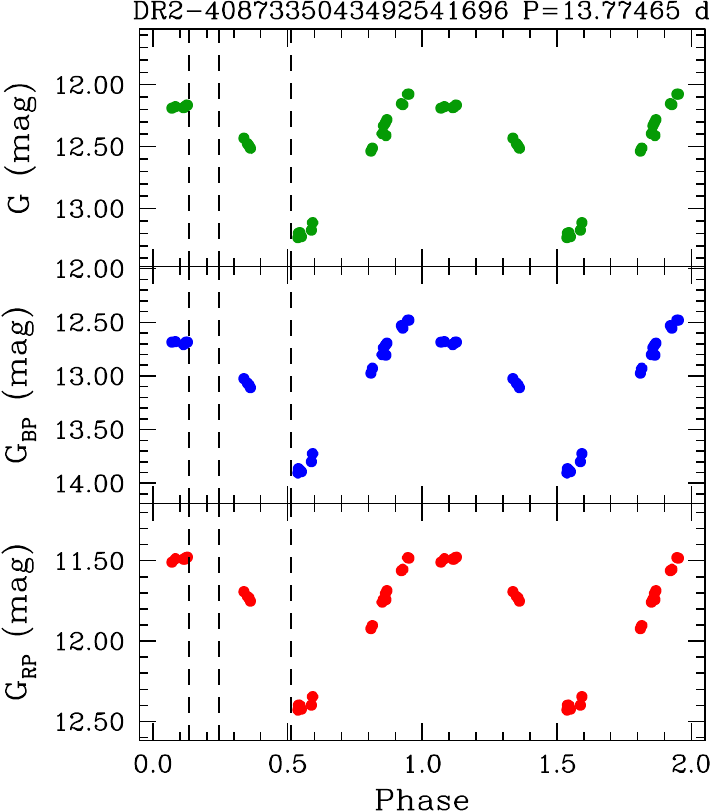}
\includegraphics[width=6.0cm]{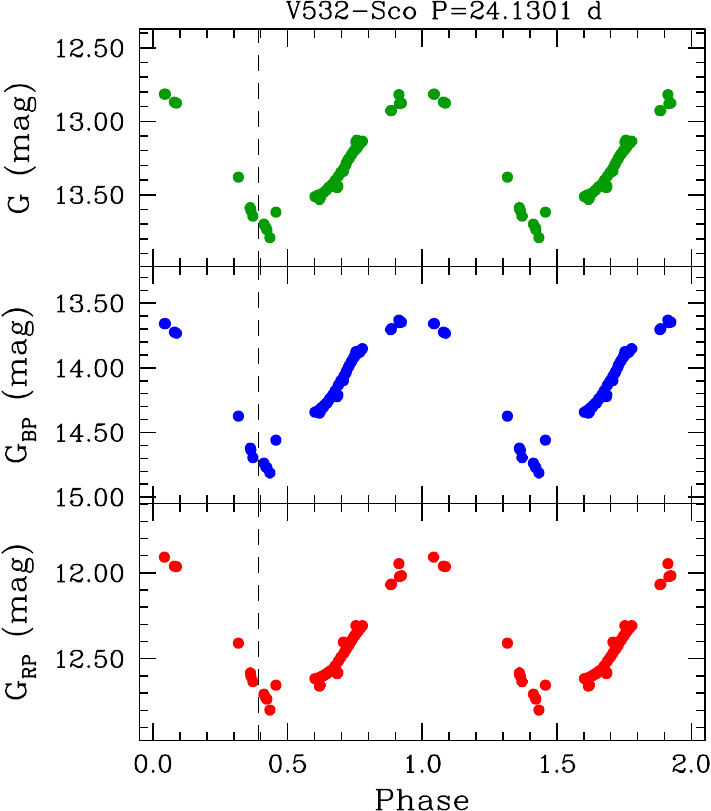}
\includegraphics[width=6.0cm]{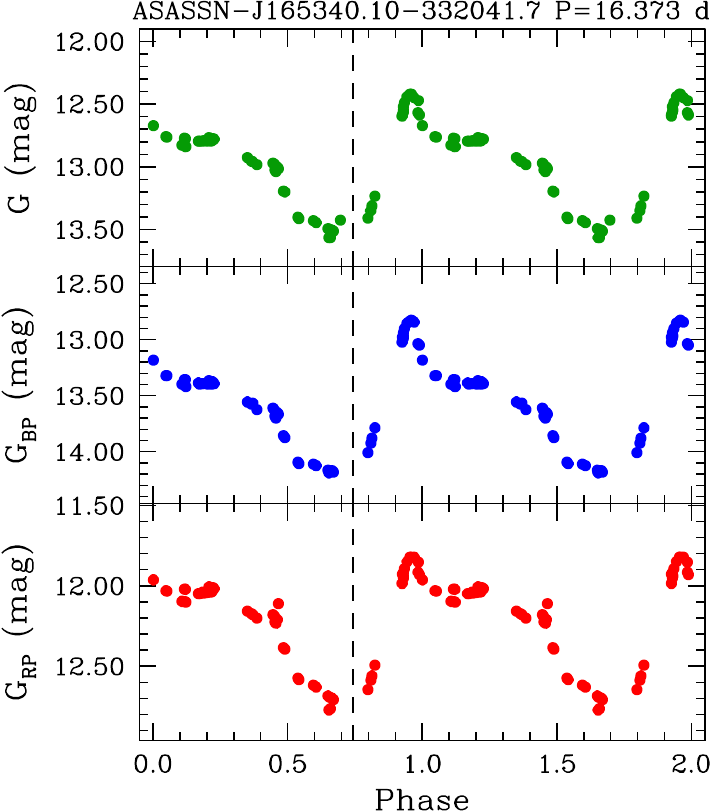}
}
\caption{Light curves in the $Gaia$ bands for the three stars with uncertain classification. The dashed lines show the phases at which the spectra were observed.}
\label{fig:lightCurves}
\end{figure*}

As mentioned in Sect.~\ref{sect:sample}, the three stars Gaia DR2 4087335043492541696, ASAS-SN J165340.10-332041.7 and V532 Sco deserve an in-depth analysis to ascertain whether or not they could be type II Cepheids. To this aim, we plot in Fig.~\ref{fig:lightCurves} the light curves in the $Gaia$ bands for the three stars, where the phases at which the spectra were observed are reported. These light curves were used by \citet{Ripepi2023} to classify these objects as fundamental DCEPs. However, the three stars have been classified as belonging to the W\,Vir class (type II Cepheids) by several authors in the literature. In more detail:

\begin{itemize}
    \item Gaia DR2 4087335043492541696, also known as V410\,Sgr or OGLE\,BLG-T2CEP-1340 has been classified as W\,Vir by several authors in the literature, e.g. \citet{Jurkovic2023} based on Kepler K2 data and \citet{Sos2020_DR2} based on $V,I$ photometry in the context of the OGLE (Optical Gravitational Lensing Experiment) survey, just to mention the most recent works. 
    \item ASAS-SN J165340.10-332041.7, also known as OGLE\,BLG-T2CEP-1089 has been considered as DCEP by \citet{Jayasinghe2018} and \citet{Skowron2019}, but lately re-classified as W\,Vir by \citet{Sos2020_DR2}.
    \item V532 Sco, also known as SV\,HV\,10484 or OGLE\,BLG-T2CEP-1229 has been classified as W\,Vir by \citet{Harris1985}, a classification subsequently confirmed by \citet{Sos2020_DR2}.   
\end{itemize}

\begin{figure*}
\centering
\includegraphics[width=15cm]{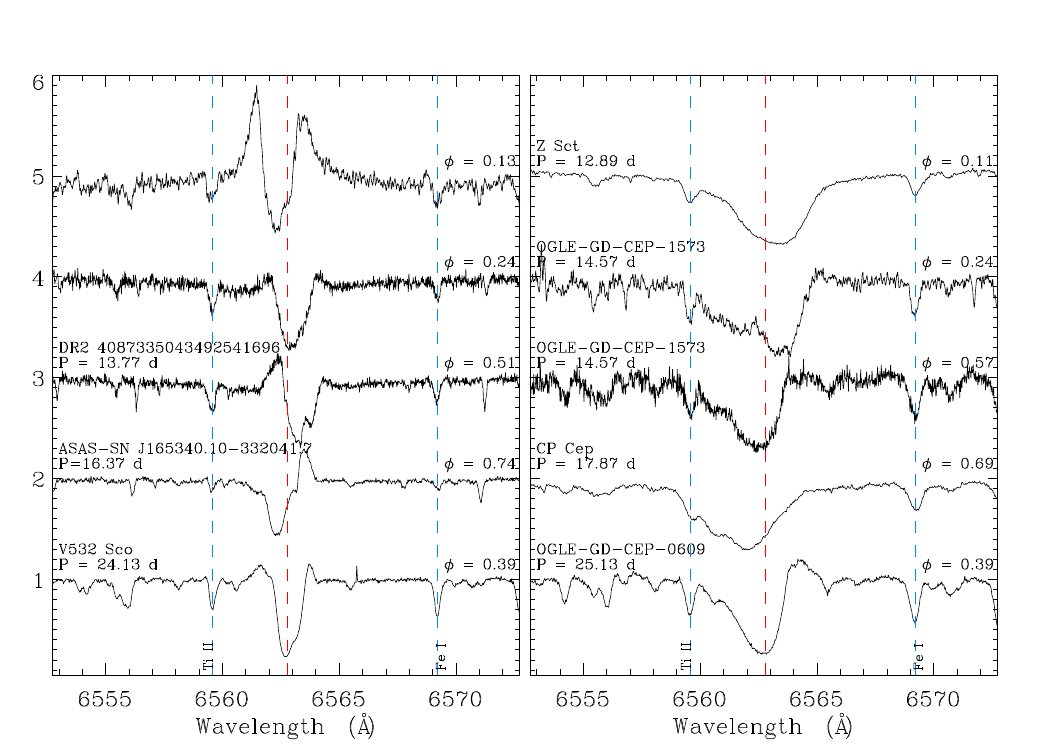}
\caption {Left panel- H$\alpha$ profiles for the three stars with uncertain classification. Profiles are plotted in the rest frame of the stellar atmosphere as measured from the metal lines. Phases at which the spectra were observed are reported in the plot. Right panel- H$\alpha$ profiles for classical cepheids extracted from our sample, with similar periods and phases.}
\label{fig:spectra_2}
\end{figure*}

We analyse now the spectra of the three stars searching for features which could possibly allow us to classify them in terms of DCEPs or W\,Vir stars. As showed in Fig.~\ref{fig:spectra_2}, we focus our attention to the $H_{\alpha}$ region where, according to \citet{Schmidt2004} two features, due to the presence of violent shocks in the moving atmospheres of the type II Cepheids can help us to distinguish between Type I and type II Cepheids: i) the difference in velocity between the $H_{\alpha}$ and the metallic lines and ii) the presence of strong emission in $H_{\alpha}$ from approximately half the cycle till phase 0.1 after the maximum. The left panel of Fig.~\ref{fig:spectra_2} shows the three epoch spectra of  Gaia DR2 4087335043492541696 and the single-epoch spectra for ASAS-SN J165340.10-332041.7 and V532 Sco. For comparison, the right panel shows the spectra of different DCEPs with periods and phases similar to those of the three investigated stars. 
The difference in velocity between $H_{\alpha}$ and the metallic lines can be clearly seen for Gaia DR2 4087335043492541696, especially at $\phi=$0.13 and 0.51, and for ASAS-SN J165340.10-332041.7. However, similar differential velocities can also be seen in DCEPs, therefore, as noted by \citet{Schmidt2004}, this feature is not a good discriminant, at least for periods larger than 8 days. However, the presence of $H_{\alpha}$ emission or P-Cygni profiles (both direct and inverse) is evident in almost all the left-panel spectra, while it is almost absent among DCEPs. An exception appears to be V532 Sco, whose spectrum is the most similar to the relative DCEPs'. 

On these bases we conclude that Gaia DR2 4087335043492541696 and ASAS-SN J165340.10-332041.7 are almost certainly W\,Vir pulsators, while the classification of V532 Sco remains uncertain. 

\begin{figure*}
\centering
\includegraphics[width=15cm]{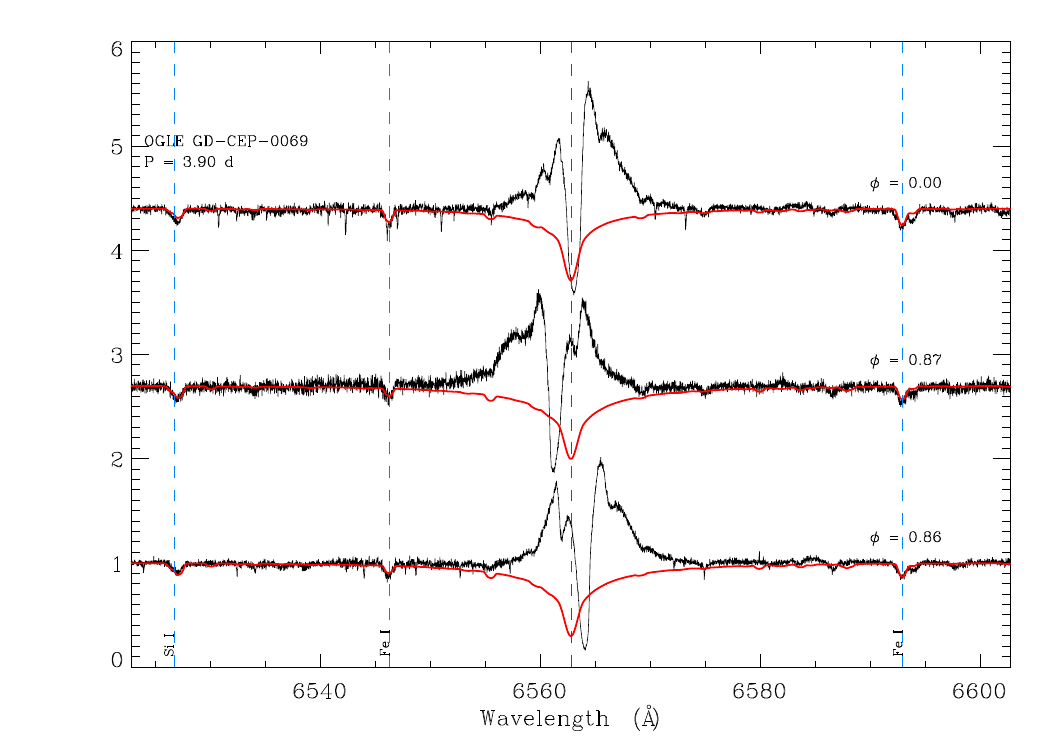}
\caption{H$\alpha$ profiles for OGLE\,CEP-CD-0069. As in Fig.~\ref{fig:spectra_2} profiles are plotted in the rest frame of the stellar atmosphere as measured from the metal lines. The overplotted synthetic spectra have been calculated with the respective parameters reported in Table~\ref{tab:logObservations} and Table~\ref{tab:abundances}. Phases at which the spectra were observed are reported in the plot.}
\label{fig:spectra_3}
\end{figure*}

As for OGLE\,CEP-GD-0069, Fig.~\ref{fig:spectra_3} show the spectra at three different phases in comparison with synthetic spectra calculated with the proper atmospheric parameters. The same arguments adopted for the previous three stars attains in this case, and we can confirm that OGLE\,CEP-GD-0069 is a type II Cepheids of BL Her subtype.

\section{Phases for seven DCEPs devoid of $Gaia$ periods and epochs of maximum light.}
\label{sec:appPhase}

To use epochs close to the spectroscopy observations we proceeded as follows: 

\begin{itemize}
    \item AP\,Sgr: the period and epoch of maximum light was calculated directly from $Gaia$ DR3 $G$-band time series using the {\tt Period04} package \citep[][]{Lenz2005}. As a result, we obtained P=5.05790 days and Epoch=58992.77858 days, where the errors are on the last digits. 
\item T\,Mon, U\,Sgr: for these stars, we have recent periods (at epoch 59591.5 days) provided by \citet{Csornyei2022}. To determine epochs close to our observations we used again {\tt Period04} but imposed the period instead of recalculating it. Therefore the adopted periods and epochs of maximum light are P=27.033911203 days; Epoch=59576.79823 days; P=6.745265182 days Epoch=59587.19128 days for T\,Mon and U\,Sgr, respectively.
\item
ASAS\,J060722+0834.0, DP\,Mon, NSVS\,2150508; V981\,Mon: for these stars we adopted periods and epochs of maximum light from the All-Sky Automated Survey for Supernovae \cite[ASAS-SN][]{Shappee2014,Christy2023}. 
\end{itemize}

\end{appendix}

\end{document}